\pdfoutput=1
\documentclass[notitlepage]{iopart}

\def \NaglowekPliku {}
\def \StopkaPliku {}

\usepackage[utf8]{inputenc}
\usepackage[OT4]{fontenc}

\usepackage{iopams}
\usepackage{enumerate}
\usepackage{graphicx}
\usepackage{calc}
\usepackage{colordvi}
\usepackage{color}
\usepackage{vmargin}
\usepackage{array}
\usepackage{subfig}

\newcommand{\bra}[1]{\left\langle#1\right|}
\newcommand{\ket}[1]{\left|#1\right\rangle}
\newcommand{\braket}[1]{\left\langle{#1}\right\rangle}
\newcommand{\melement}[3]{\braket{#1\left|#2\right|#3}}

\newcommand{\gr}{{\rm gr}}
\newcommand{\scl}{{\rm sc}}
\newcommand{\tot}{{\rm tot}}
\newcommand{\phys}{{\rm phys}}
\newcommand{\kin}{{\rm kin}}
\newcommand{\KKL}{\mathrm{KKL}}
\newcommand{\AL}{\mathrm{AL}}
\newcommand{\DC}{\mathrm{DC}}
\newcommand{\LQG}{\mathrm{LQG}}
\newcommand{\diff}{\mathrm{diff}}
\newcommand{\inn}{\mathrm{in}}
\newcommand{\out}{\mathrm{out}}

\renewcommand{\tr}{\mathrm{Tr}}
\newcommand{\inv}{\mathrm{Inv}}

\newcommand{\id}{\idd}	
	\newcommand{\idd}{\mathbf{1}}
	\newcommand{\iddd}{\mathbb{1}}
	\newcommand{\idddd}{\mathbb{I}}
\newcommand{\im}{\dot\imath}
\newcommand{\dd}{\mathrm{d}}
\newcommand{\pd}{\partial}

\newcommand{\K}[1]{\mathbb{#1}}	
\newcommand{\T}[1]{\mathcal{#1}}	

\newcommand{\N}{\T{N}}	

\newcommand{\Hil}{\mathcal{H}}
\newcommand{\Hal}{\Hil_{\mathrm{AL}}}
\newcommand{\Hd}{\Hil_{\mathrm{DC}}}
\newcommand{\Hgr}{{\Hil_{\gr}}}
\newcommand{\Hsc}{{\Hil_{\scl}}}
\newcommand{\Htot}{{\Hil_{\tot}}}
\newcommand{\Hgrb}{{\Hil_{\gr,B}}}
\newcommand{\Hphys}{{\Hil_{\phys}}}

\newcommand{\Span}{{\rm Span}}
\newcommand{\spann}[2]{\begin{array}{c}span\\ \;_{\substack{#1}}\end{array} \left(#2\right)}

\newcommand{\Cgr}{{\widehat{C_{\gr}}}}
\newcommand{\Csc}{{\hat{C}_{\scl}}}
\newcommand{\Ctot}{{\hat{C}_{\tot}}}

\newcommand{\tab}{\qquad\qquad}
\newcommand{\taab}{\tab\tab}

\newcommand{\omicron}{{\cal I}}
\newcommand{\ppi}{{\scriptstyle\Pi}}
\newcommand{\F}{{\mathcal F}}

\newcommand{\mnote}[1]{{\Red{*}}\marginpar{%
      \vskip-\baselineskip
      \raggedright\footnotesize
      \itshape\hrule\smallskip\tiny{\Red{#1}}\par\smallskip\hrule}}

\newcommand{\reef}[1]{(\ref{eq:#1})} 
\renewcommand{\fref}[1]{figure~\ref{fig:#1}} 
\renewcommand{\sref}[1]{section~\ref{sc:#1}} 
\renewcommand{\Fref}[1]{Figure~\ref{fig:#1}} 
\renewcommand{\Sref}[1]{Section~\ref{sc:#1}} 

\def\be{\begin{equation}}
\def\ee{\end{equation}}
\def\ba{\begin{eqnarray}}
\def\ea{\end{eqnarray}}
\def\lp{{\ell}_{\rm Pl}}
\def\g{\gamma}

\newcommand{\s}[1]{{#1}^{(s)}}

\begin{document}

\ifx \NaglowekPliku \undefined
  \def \NaglowekPliku {
\documentclass[notitlepage]{iopart}
\usepackage[utf8]{inputenc}
\usepackage[OT4]{fontenc}

\usepackage{iopams}
\usepackage{enumerate}
\usepackage{graphicx}
\usepackage{calc}
\usepackage{colordvi}
\usepackage{color}
\usepackage{vmargin}
\usepackage{array}
\usepackage{subfig}




\newcommand{\bra}[1]{\left\langle#1\right|}
\newcommand{\ket}[1]{\left|#1\right\rangle}
\newcommand{\braket}[1]{\left\langle{#1}\right\rangle}
\newcommand{\melement}[3]{\braket{#1\left|#2\right|#3}}

\newcommand{\gr}{{\rm gr}}
\newcommand{\scl}{{\rm sc}}
\newcommand{\tot}{{\rm tot}}
\newcommand{\phys}{{\rm phys}}
\newcommand{\kin}{{\rm kin}}
\newcommand{\KKL}{\mathrm{KKL}}
\newcommand{\AL}{\mathrm{AL}}
\newcommand{\DC}{\mathrm{DC}}
\newcommand{\LQG}{\mathrm{LQG}}
\newcommand{\diff}{\mathrm{diff}}
\newcommand{\inn}{\mathrm{in}}
\newcommand{\out}{\mathrm{out}}

\renewcommand{\tr}{\mathrm{Tr}}
\newcommand{\inv}{\mathrm{Inv}}

\newcommand{\id}{\idd}	
	\newcommand{\idd}{\mathbf{1}}
	\newcommand{\iddd}{\mathbb{1}}
	\newcommand{\idddd}{\mathbb{I}}
\newcommand{\im}{\dot\imath}
\newcommand{\dd}{\mathrm{d}}
\newcommand{\pd}{\partial}

\newcommand{\K}[1]{\mathbb{#1}}	
\newcommand{\T}[1]{\mathcal{#1}}	

\newcommand{\N}{\T{N}}	

\newcommand{\Hil}{\mathcal{H}}
\newcommand{\Hal}{\Hil_{\mathrm{AL}}}
\newcommand{\Hd}{\Hil_{\mathrm{DC}}}
\newcommand{\Hgr}{{\Hil_{\gr}}}
\newcommand{\Hsc}{{\Hil_{\scl}}}
\newcommand{\Htot}{{\Hil_{\tot}}}
\newcommand{\Hgrb}{{\Hil_{\gr,B}}}
\newcommand{\Hphys}{{\Hil_{\phys}}}

\newcommand{\Span}{{\rm Span}}
\newcommand{\spann}[2]{\begin{array}{c}span\\ \;_{\substack{#1}}\end{array} \left(#2\right)}

\newcommand{\Cgr}{{\widehat{C_{\gr}}}}
\newcommand{\Csc}{{\hat{C}_{\scl}}}
\newcommand{\Ctot}{{\hat{C}_{\tot}}}

\newcommand{\tab}{\qquad\qquad}
\newcommand{\taab}{\tab\tab}

\newcommand{\omicron}{{\cal I}}
\newcommand{\ppi}{{\scriptstyle\Pi}}
\newcommand{\F}{{\mathcal F}}


\newtheorem{lm}{Lemma}
\newtheorem{thm}[lm]{Theorem}
\newtheorem{df}{Definition}

\newcommand{\mnote}[1]{{\Red{*}}\marginpar{%
      \vskip-\baselineskip
      \raggedright\footnotesize
      \itshape\hrule\smallskip\tiny{\Red{#1}}\par\smallskip\hrule}}

\newcommand{\reef}[1]{(\ref{eq:#1})} 
\renewcommand{\fref}[1]{figure~\ref{fig:#1}} 
\renewcommand{\sref}[1]{section~\ref{sc:#1}} 
\renewcommand{\Fref}[1]{Figure~\ref{fig:#1}} 
\renewcommand{\Sref}[1]{Section~\ref{sc:#1}} 


\def\be{\begin{equation}}
\def\ee{\end{equation}}
\def\ba{\begin{eqnarray}}
\def\ea{\end{eqnarray}}
\def\lp{{\ell}_{\rm Pl}}
\def\g{\gamma}

\newcommand{\s}[1]{{#1}^{(s)}}

\begin{document}
}
\fi
\ifx \StopkaPliku \undefined
  \def \StopkaPliku {\end{document}}
\fi
\NaglowekPliku


\title{One vertex spin-foams with the Dipole Cosmology boundary}
\author{Marcin Kisielowski${}^{1,2}$, Jerzy Lewandowski${}^{1}$ and Jacek Puchta${}^{1,3}$}

\address{${}^1$ Instytut Fizyki Teoretycznej, Uniwersytet Warszawski,
ul. Ho{\.z}a 69, 00-681 Warszawa (Warsaw), Polska (Poland)}
\address{${}^2$ St. Petersburg Department of Steklov Mathematical Institute, Russian Academy of Sciences,
Fontanka 27, St. Petersburg, Russia}
\address{${}^3$ Centre de Physique Theorique de Luminy,
Case 907, Luminy, F-13288 Marseille, France}

\eads{\mailto{lewand@fuw.edu.pl}, \mailto{cino@fuw.edu.pl}, \mailto{jpa@fuw.edu.pl}}

\begin{abstract}
We find all the spin-foams contributing in the first order of the vertex expansion to the transition amplitude of the  Bianchi-Rovelli-Vidotto Dipole  Cosmology model. Our algorithm is general and provides  spin-foams of arbitrarily given, fixed:  boundary and, respectively, a number of internal vertices. We use the recently introduced  Operator Spin-Network Diagrams framework.  
\end{abstract}

\StopkaPliku

\ifx \NaglowekPliku \undefined 
  \def \NaglowekPliku {
\documentclass[notitlepage]{iopart}
\begin{document}
}
\fi
\ifx \StopkaPliku \undefined
  \def \StopkaPliku {\end{document}}
\fi
\NaglowekPliku

\section{Introduction}\label{sc:OSN_diagrams}
Spin-foams are quantum histories of states of the gravitational field according to the Spin-Foam Models of quantum gravity. In the usual formulation, a spin-foam is a two-complex, whose faces are colored with  representations of a given group (depending on a model, for example SU(2)) and edges are colored with invariants of the tensor products \cite{RR,Baezintro, perez, Rovellibook}, or equivalently with operators if one uses the Operator Spin-Foam framework \cite{Operator_SF}. The spin-foams encode the data necessary to calculate the transition amplitude between states of Loop Quantum Gravity \cite{Rovellibook, AshLewrev, Marev, Ashtekarbook, Thiemannbook, LQGdiscr} or more generally,  the Rovelli boundary transition amplitude \cite{Rovellibook}. There are a few candidates for the spin-foam model of Quantum Gravity \cite{BC,EPRL,flipped,FK,HT,You,E}. Important for the compatibility with Loop Quantum Gravity is to admit sufficiently general class of the 2-complexes, such that all the (closed, either abstract or embedded in a 3-manifold) graphs are obtained as their boundaries \cite{SFLQG}. The optimal  class of such 2-cell complexes was proposed in \cite{OSD}. They are naturally provided in terms of the diagrammatic formalism introduced therein and called operator spin-network diagrams (OSN diagrams). A similar diagramatic framework for triangulations was introduced before in \cite{FrankPhD}. An additional advantage of our formalism,  is that the OSN diagrams do not require neither 3d nor 4d imagination, they are easy to use and to classify possible spin-foams. We utilize and even improve these technical advantages in the current work.

The generalized (to a non-simplicial 2-cell complex) EPRL vertex has been recently applied to introduce Dipole Cosmology, a quantum cosmological model which opens a new theory that can be called Spin-Foam Cosmology \cite{SF_cosmology}. This application of spin foams in cosmology gave us the motivation to do the current research. We apply here the operator spin-network diagrams framework to find  all spin-foams which contribute to the boundary amplitude of a fixed spin-network state in given order of the vertex expansion. The technical task one encounters when solving this problem is finding all the diagrams whose boundary is a given graph. We solve this problem in \sref{piany_o_ustalonym_brzegu} with the use of squid sets we introduce in \sref{oriented_squid_sets}. The solution we present in that section is not limited to the Dipole  Cosmology model only. Actually, it applies to the general spin-foam case. In \sref{dipole_cosmology} we apply this general scheme to the model of Dipole Cosmology \cite{SF_cosmology}. We find all the OSN-diagrams whose boundary is the boundary graph of Dipole Cosmology and which have exactly one interaction graph. Those diagrams contribute to the boundary amplitude in the first order of vertex expansion.

\subsection{Definition of OSN-diagrams}\label{def}
In this subsection we recall the definition of OSN diagrams we introduced in \cite{OSD}. In the analogy to spin-foams, which are colored 2-complexes, OSN-diagrams are colored graph diagrams. We first recall the definition of graph diagrams and then we recall the definition of coloring which turns a graph diagram into an OSN-diagram.

One may think of graph diagrams as a way of building 2-complexes from building blocks which are (suitable) neighborhoods of vertices of the corresponding foam \cite{RR,SFLQG}. Such a neighborhood is a 2-complexes obtained as an image of a homotopy of a graph. When glued together, they form a 2-complex. The way one glues them together is encoded in certain relations.

Strictly speaking a graph diagram $(\T G,\T R)$ consists of a set ${\T G}$ of oriented, connected, closed graphs $\{\Gamma_1, ... ,\Gamma_N\}$ and a family $\T R$ of relations defined as follows (see \fref{1_osn_diagram}):
\begin{figure}[ht!]
  \centering
\includegraphics[width=0.6\textwidth]{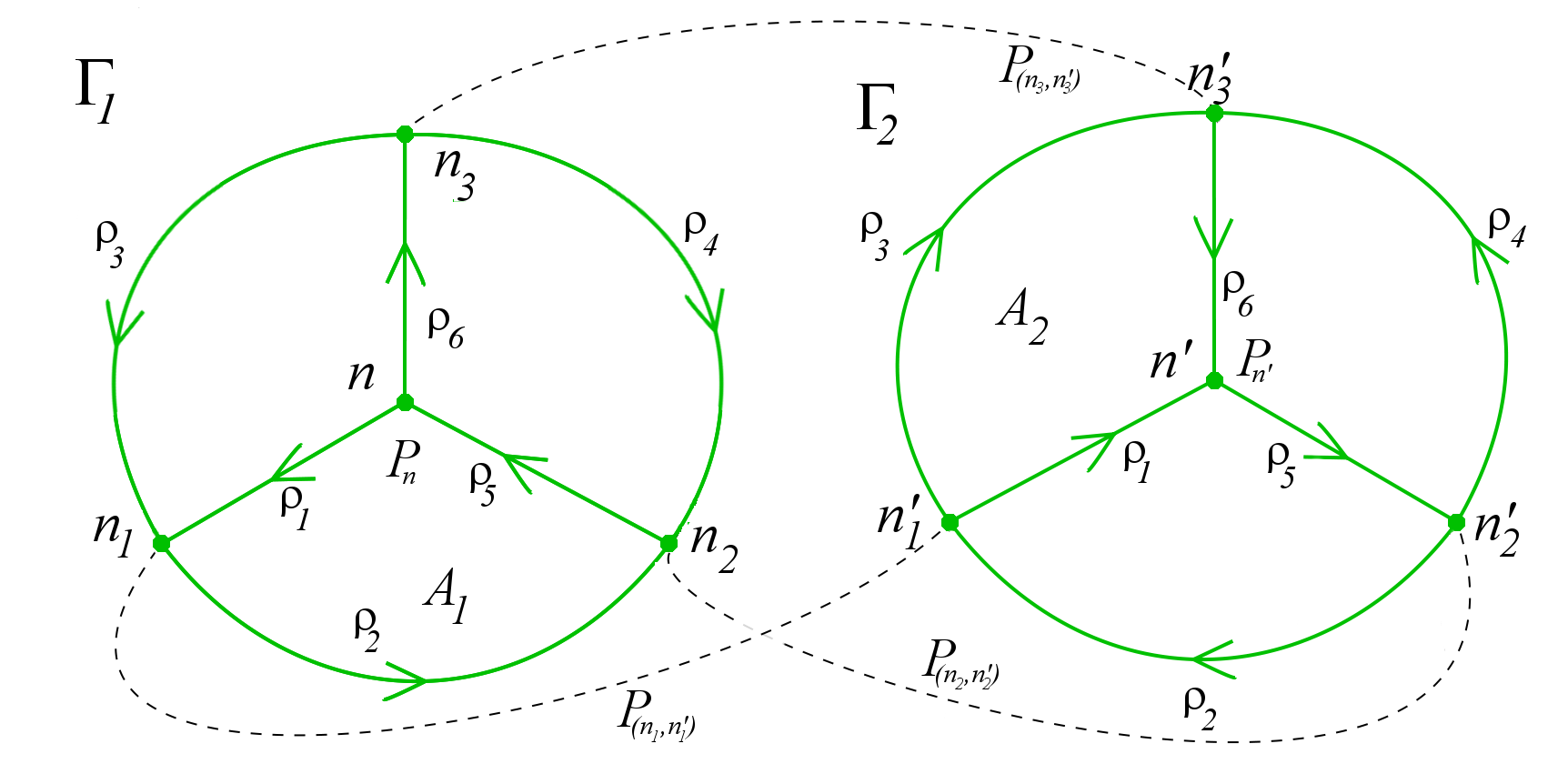}
  \caption{An operator spin-network diagram. The dashed curves mark the node relation. The link relation (unmarked) relates each link at $n_I$, $I=1,2,3$  with the link at $n'_I$ colored by  $\rho_K$ of a same $K$.  Operators $P_{n_In'_I}$, $I=1,2,3$ mark the pairs of related nodes, whereas the operators $P_n$ and $P_{n'}$ mark the unrelated nodes $n$ and, respectively, $n'$. Each connected component $\Gamma_1$ and $\Gamma_2$ of the green graph is colored by contractors $A_1$ and $A_2$, respectively.}
  \label{fig:1_osn_diagram}
\end{figure}

\begin{itemize}
	\item $\T R_{\rm node}$: a symmetric relation in the set of nodes of the graphs which we call the node relation, such that each node $n$ is either in relation with precisely one $n'\not=n$ or is unrelated (in the later case, it is called a boundary node).

	\item $\T R_{\rm link}$: a family of symmetric relations in the set of links of the graphs which we call collectively the link relation. If a node $n$ is in relation with a node $n'$, then we define a bijective map between incoming/outgoing links at $n$, with  outgoing/incoming links at $n'$; no link is left free neither at the node $n$ nor at $n'$; two links identified with each other by the bijection are called to be in the relation $\T R^{(n,n')}_{\rm link}$ at the pair of nodes $n,n'$; a link which intersects $n$/$n'$ twice,  emerges in the relation twice: once as an incoming and once as an outgoing link.
\end{itemize}

In order to be related, two nodes have to satisfy the consistency condition: the number of the incoming/outgoing links at each of them has to coincide with the number of the outgoing/incoming links at the other one (with possible closed links counted twice). Note that two graphs can be treated as one disconnected graph. Thus to reduce that ambiguity we assume that all the graphs defining the diagram are connected.
 
An operator spin-network diagram $({\cal G}=\{\Gamma_1,...,\Gamma_N\}, \T R, \rho, P, A)$ is defined by using a compact group $G$ and coloring a graph diagram $({\cal G},\T R)$ as follows (see \fref{1_osn_diagram}):
\begin{itemize}
	\item The coloring $\rho$ assigns to each link $\ell$ of each graph $\Gamma_I$, $I=1,...,N$ an irreducible representation of the group $G$: \be \ell\mapsto\rho_\ell. \ee It is assumed that whenever two links $\ell$ and $\ell'$ are mapped to each other by the link relation $\T R_{\rm link}^{nn'}$ at some nodes $n$ and $n'$, then \be\label{eq:RhoSaRowne} \rho_\ell= \rho_{\ell'}.\ee
	\item The coloring $P$ assigns to each node $n$ an operator: \be\label{eq:nodeoperator} n\mapsto P_n\in \Hil_{n}\otimes \Hil_{n}^*, \ee where $\Hil_n$ is a Hilbert space defined at each node in the following way:
\be\label{eq:H.n}
	\Hil_n\ =\ \inv\left(\bigotimes_i \Hil_{\rho_i}^* \otimes \bigotimes_j\Hil_{\rho_j}\right)\;\subset\;\left(\bigotimes_i \Hil_{\rho_i}^* \otimes \bigotimes_j\Hil_{\rho_j}\right)
\ee
where $i$/$j$ labels the links incoming/outgoing at $n$.

Whenever two nodes $n$ and $n'$ are related by $\T R_{\rm node}$, then (from \reef{RhoSaRowne} and \reef{H.n}) it follows that $\Hil_n=\Hil_{n'}^*$ and  it is assumed about $P$ that \be\label{eq:DualityOfPs} P_n=P_{n'}^*\ee
	\item The coloring $A$ assigns to each graph $\Gamma_I$ a tensor, which we call contractor:
		\be\label{eq:A.Gamma}
			\Gamma_I\mapsto A_\Gamma\in\ \left(\bigotimes_n {\cal H}_n\right)^*
		\ee
		where $n$ runs through the nodes of $\Gamma_I$.
\end{itemize}

Each graph $\Gamma_I$ itself defines a contractor, in the sense that there is a natural contraction defined by the graph $\Gamma_I$ and by the natural trace operation in $\bigotimes_{\ell}{\cal H}_\ell\otimes{\cal H}_\ell^*$ which contains $\bigotimes_n {\cal H}_n$, where $n$~/~$\ell$ ranges the set of nodes/links of $\Gamma_I$. We denote this natural contractor by $A^{\rm Tr}_{\Gamma_I}$.  However,  that natural contraction is often preceded by some additional operations, like the EPRL embedding which gives rise to the EPRL operator spin-network diagrams.

\StopkaPliku

\ifx \NaglowekPliku \undefined
  \def \NaglowekPliku {
\documentclass[notitlepage]{iopart}
\begin{document}
}
\fi
\ifx \StopkaPliku \undefined
  \def \StopkaPliku {\end{document}}
\fi
\NaglowekPliku


\section{Characterization and construction of the diagrams}\label{sc:oriented_squid_sets}
We will introduce now a useful characterization of the diagrams. The characterization will allow  to control the structure and the complexity of the diagrams in a clear way. Finally it will lead to algorithms  for construction of diagrams. In this section we will introduce the first, simpler algorithm. It will be improved to produce diagrams of a given boundary, in the next section.
   
\subsection{Squids}
An important element of our characterization of  graph diagrams is  an {\it oriented squid set}. Given a graph (oriented and closed), its oriented squid set is obtained by removing from each link of the graph a point of its entry and shaking the whole thing so that the disconnected parts of each link {\it go apart} (\fref{2_01_g2oss}).

\begin{figure}[ht!] 
	\centering
	\includegraphics[width=0.60\textwidth]{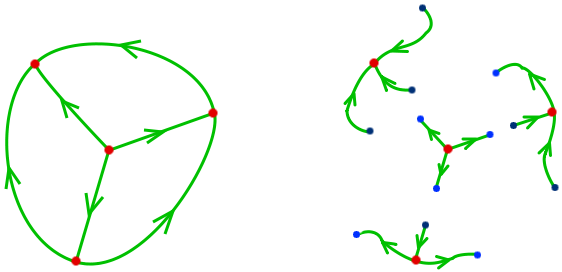}
	\caption{A graph (on the LHS)  and its oriented squid set (on the RHS) which consists of four 3-leg squids.}
	\label{fig:2_01_g2oss}
\end{figure}

Two different graphs may define a single squid set. Therefore, it makes sense to introduce and consider the notion of a squid set on its own. An {\it oriented squid} consist of $1$ point called {\it head}, $k^+$ legs beginning at the head, and $k^-$ legs ending at the head. In other words it may be considered as the topological space obtained by glueing $k^++k^{-}$ oriented intervals with the head in the suitable way (\fref{2_02_o_squid}).
\begin{figure}[ht!]
	\centering
	\subfloat[$\;$]
		{\includegraphics[width=0.32\textwidth]{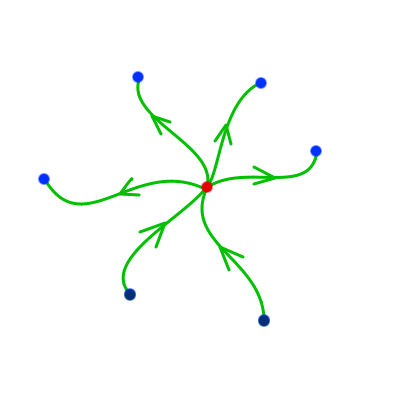}\label{fig:2_02_o_squid}}
	\hspace{0.1\textwidth}
	\subfloat[$\;$]
		{\includegraphics[width=0.42\textwidth]{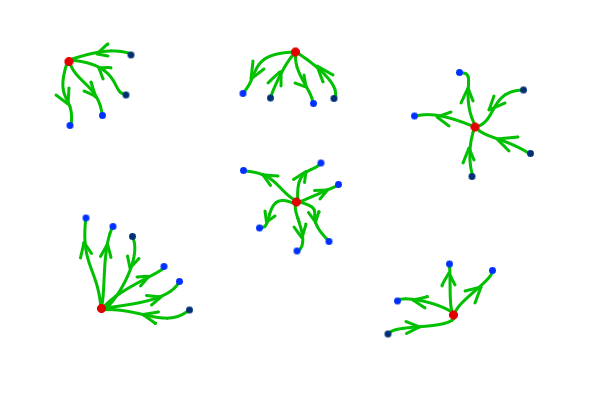} \label{fig:2_03_oss}}
	\caption{(a) The oriented squid of $k^-=2$ incoming,  and $k^+=4$ outgoing 
	legs. (b) An example of an oriented squid set.}
	\label{fig:2_squid}
\end{figure}
An {\it oriented squid set}  ${\cal S}$ is a disjoint finite union of squids (\fref{2_03_oss}).

A graph $\Gamma$ consists of a squid set ${\cal S}_{\Gamma}$ and information about glueing the legs of the squids. Conversely, given a squid set ${\cal S}$, a  graph  may be obtained by glueing the end of each outgoing leg with the beginning of arbitrarily chosen incoming leg of either the same or a different squid. However this procedure neither is unique, nor there is a guarantee it can be completed. Therefore, a single squid set can be the squid set of either more then one graph, or of an exactly one graph, or of no graph at all (\fref{2_04_oss_amb}).
\begin{figure}[ht!]
	\centering
	\subfloat[$\;$]
		{\includegraphics[width=0.40\textwidth]{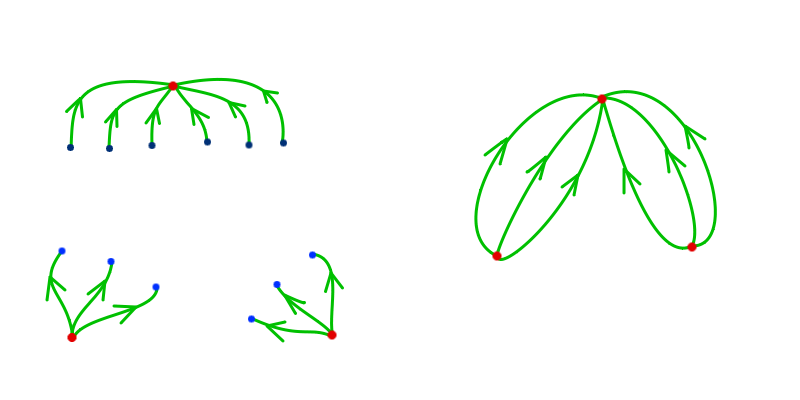} }
	\hspace{0.1\textwidth}
	\subfloat[$\;$]
		{\includegraphics[width=0.40\textwidth]{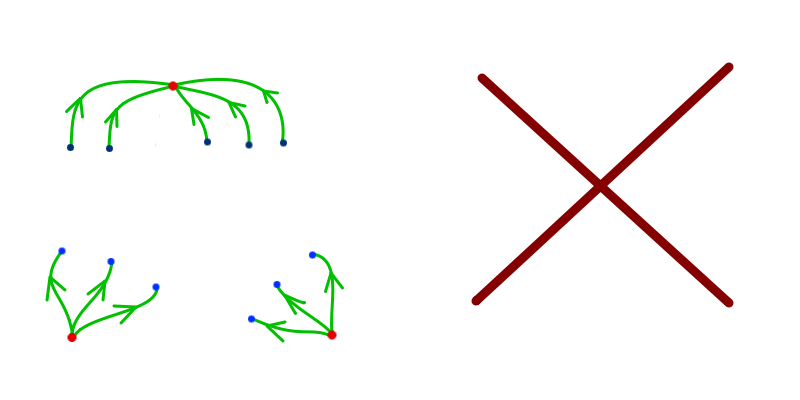} }
	\\
	\subfloat[$\;$]
		{\includegraphics[width=0.40\textwidth]{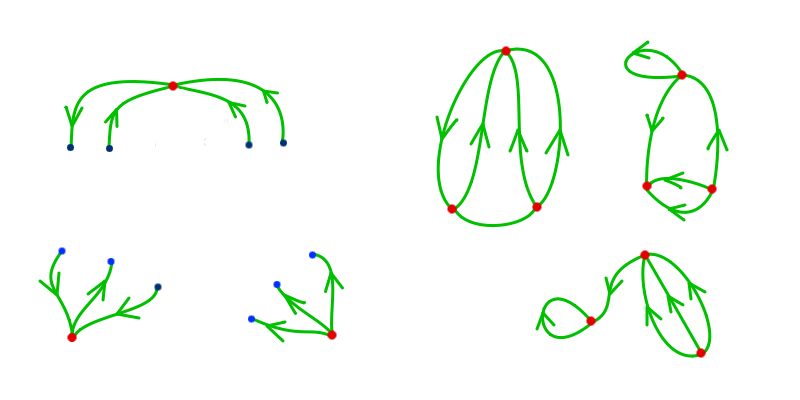} }
	\caption{(a) An oriented squid set (LHS) and a unique oriented graph obtained by glueing the legs (RHS). (b) An oriented squid set whose legs can not be glued to give an oriented graph. (c) An oriented squid set (LHS), and three different oriented graphs (RHS) each of which can be obtained by glueing the legs in a suitable way.}
	\label{fig:2_04_oss_amb}
\end{figure}

A graph diagram $(\Gamma,{\cal R})$ can also be obtained from a squid set   ${\cal S}_\Gamma$  by endowing it  with additional structure. The part of the structure has been already described above, it allows to reconstruct the graph $\Gamma$.  The second element of the structure, the node relation ${\cal R}^{\rm node}$,  is given by indicating a set of pairs $\{\{\lambda_{1},\lambda'_{1}\}, ..., \{\lambda_{N},\lambda'_{N}\}\}$ of squids $\lambda_{i},\lambda'_{i}\in {\cal S}$, whose heads are in the node relation ${\cal R}^{\rm node}$.  The last element of the graph diagram structure, a link relation,  is defined for each pair $\{\lambda_{i},\lambda'_{i}\}$ as a bijection between  the incoming/outgoing legs of the squid  $\lambda_{i}$ with the outgoing/incoming legs of the squid $\lambda'_{i}$ (\fref{2_05_relations}).
\begin{figure}[ht!]
	\centering
	\includegraphics[width=0.50\textwidth]{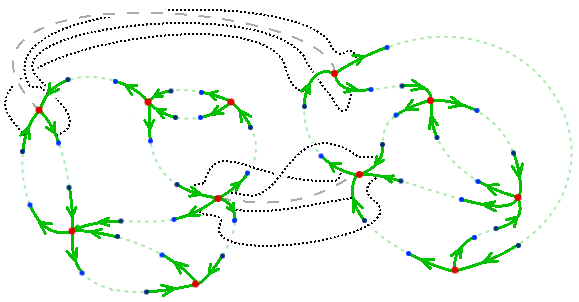}
	\caption{The construction of a graph diagram from a squid set: $(i)$ the solid curves are squid legs meeting at squid heads $(ii)$ the dashed curves connecting the ends of the squid legs define glueing the legs into a (set of connected) graph(s), $(iii)$ the dashed curves connecting the squid heads define a node relation, $(iv)$ the dotted curves define a link relation at each pair of related nodes.}
	\label{fig:2_05_relations}
\end{figure}

\subsection{The algorithm}\label{sc:algorithm1}
This characterization of an arbitrary graph diagram leads to the following algorithm for construction of all the graph diagrams:

\begin{enumerate}
\item Squids: fix a squid set ${\cal S}$ (\fref{2_06_diagram_1_s})
\begin{figure}[ht!]
	\centering
	\includegraphics[width=0.60\textwidth]{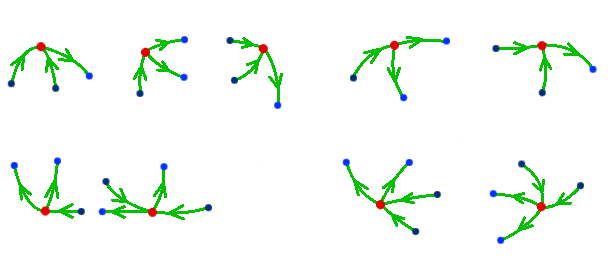}
	\caption{An oriented squid set $\T S$.}
	\label{fig:2_06_diagram_1_s}
\end{figure}
\item Node relation:  choose $N$ pairs $\lambda_{i},\lambda'_i$, $i=1,...,N$ of consistent squids (the numbers of incoming/outgoing legs of unprimed squid in each pair coincides with the numbers of the outgoing/incoming legs in the primed squid); an element $\lambda\in{\cal S}$ can emerge only once in this set of pairs or not at all (\fref{2_07_diagram_2_n}). The relation ${\cal R}_{\rm node}$ is  such that the heads of the chosen squids $\lambda_i$ and $\lambda'_i$, for $i=1,...,N$ are related to each other, and to nothing else.
\begin{figure}[ht!]
	\centering
	\includegraphics[width=0.60\textwidth]{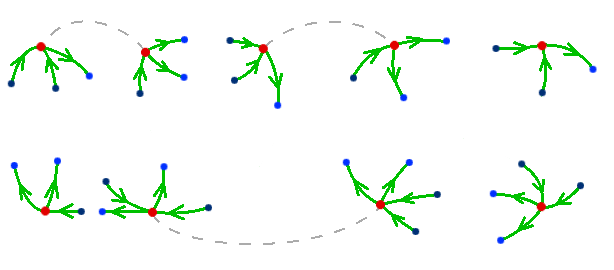}
	\caption{Three pairs of squids have been chosen to be related by a node relation.}
	\label{fig:2_07_diagram_2_n}
\end{figure}

\item Link relation: for each pair $\lambda_i,\lambda'_i$ of the squids, $i=1,...,N$, define a bijective map carrying the incoming/outgoing legs on one squid into the outgoing/incoming legs of the other squid.   

\item Glueing: glue the end of each outgoing leg with the beginning of exactly one incoming leg of either the same or a different squid (\fref{2_08_diagram_3_g}).
\begin{figure}[ht!]
	\centering
	\includegraphics[width=0.6\textwidth]{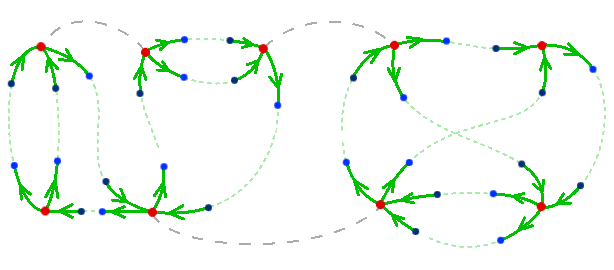}
	\caption{One of the graph diagrams that can be constructed from $\T S$ (the link relation is omitted).}
	\label{fig:2_08_diagram_3_g}
\end{figure}

\item Given the data $(i) - (iii)$ perform all possible options of the glueing $(iv)$ - see \fref{2_09_diagram_4}. 

\end{enumerate}
\begin{figure}[ht!]
	\centering
\begin{tabular}{>{\centering}p{0.5\textwidth}>{\centering}p{0.5\textwidth}}
	\subfloat[$\;$]{\includegraphics[width=0.45\textwidth]{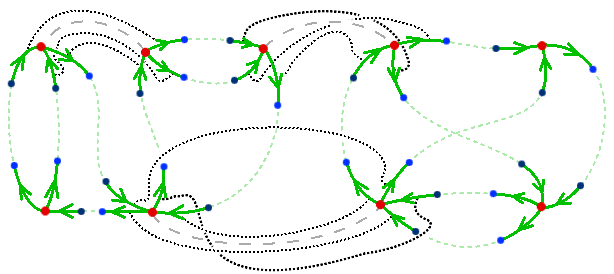} }&
	\subfloat[$\;$]{\includegraphics[width=0.45\textwidth]{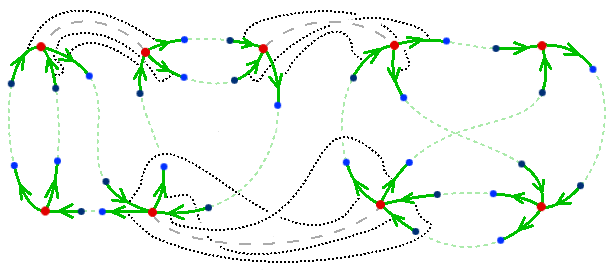} }\tabularnewline
	\multicolumn{2}{>{\centering}p{\textwidth}}{\subfloat[$\;$]{\includegraphics[width=0.45\textwidth]{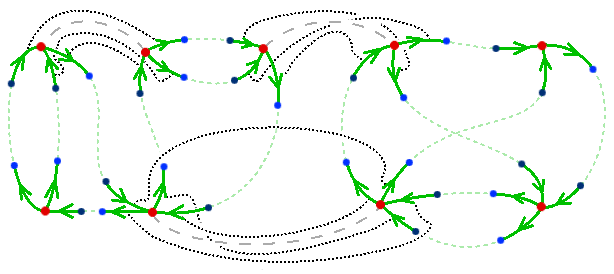} }}
\end{tabular}
	\caption{Three different graph diagrams obtained by different choices of link relation in  \fref{2_08_diagram_3_g}. In total, there are eight different graph diagrams for this choice of a node relation and many more for arbitrary choice of a node relation.
	}
	\label{fig:2_09_diagram_4}
\end{figure}

A  graph diagram ${\cal D}$ resulting from $(i)-(iv)$ consist of a graph 
$$\Gamma\ =\ \Gamma_1\cup...\cup\Gamma_N,$$ 
the disjoint union of connected graphs $\Gamma_I$, $I=1,..,N$ obtained by the glueing $(iv)$, a node relation given by $(ii)$ and a link relation provided by $(iii)$. 
We denote it shortly by  ${\cal D}=(\{\Gamma_1,...,\Gamma_N\}, {\cal R})$. 
It is turned into an operator spin-network diagram by coloring the links, the nodes and the connected components of $\Gamma$ according to Section \ref{def}.

\subsection{Discussion}
\subsubsection{Colorings}

The colorings may be constrained by the geometry of the graph diagram and the relations. In order to control this constraint one may from the beginning fix the coloring of the boundary links (that is the legs of the squids whose heads are boundary nodes - the nodes which are unrelated by the node relation)  by representations, and then allow only such glueings that agree with the coloring (i.e. two legs may be glued if and only if the are colored by the same representation).

\subsubsection{Reorientation of the diagrams}

Graphs used in the definition of  OSN diagram are oriented, and the orientation is relevant for the node and link relations. However, the operators evaluated from the OSN diagram are invariant with respect to consistent changes of the orientation accompanied with the dualization of the representation colors.

Suppose $(\{\Gamma_1',...,\Gamma_N'\}, {\cal R}',\rho',P',A')$ is an OSN-diagram  obtained from a given OSN-diagram $(\{\Gamma_1,...,\Gamma_N\}, {\cal R},\rho,P,A)$ by: flipping the orientation of some of the links $\ell_1,...\ell_k$, leaving the same node relations, and leaving the same link relations, setting
\begin{equation}\rho'_{\ell_i^{-1}}\ =\ \rho^*_{\ell_i}, \ \ \ i=1,...,k,\end{equation}
leaving 
\begin{equation}\rho'(\ell)=\rho(\ell)\end{equation} 
for each unflipped link $\ell$, and $P'=P$ as well as $A'=A$. Notice, that the transformation of the orientations and the labelling $\rho\mapsto\rho'$ preserves the Hilbert spaces ${\cal H}_n$, where $n$ ranges the set of the nodes of the diagram graph. That property makes the choice $P'=P$ and $A'=A$ possible. The OSN-diagram $(\{\Gamma_1',...,\Gamma_N'\}, {\cal R}',\rho',P',A')$ can be considered as {\it reoriented} OSN-diagram $(\{\Gamma_1,...,\Gamma_N\}, {\cal R},\rho,P,A)$. The reorientation of any OSN-diagram does not change the Hilbert spaces and the resulting operator.    

Notice however, that given an OSN-diagram $(\{\Gamma_1,...,\Gamma_N\}, {\cal R},\rho,P,A)$, we are not free to reorient any link we want, say exactly one arbitrarily selected  link, to obtain a reoriented OSN-diagram  $(\{\Gamma_1',...,\Gamma_N'\}, {\cal R}')$. The possible changes of the orientation have to be consistent with the node-link relations ${\cal R}$. For each of the flipped links $\ell_i$, a link which is in the link relation with $\ell_i$ at one of its nodes has to be flipped too, and so on. In \cite{OSD} we identified the chains of links related to each other with the equivalence classes of suitably defined face relation. We characterised the face relation classes in detail. It follows, that for each of the flipped links $\ell_i$, all the links, elements of its face relation equivalence class have to be flipped as well. This is the necessary and sufficient consistency condition for reorienting the links of a graph diagram in a way consistent with the node-link relations.

\subsubsection{Advantages and a drawback} 

The characterization and construction presented above allows to control the complexity of the diagrams by the following measures: the number of internal edges, the number of disconnected components of the graphs, the complexity of each disconnected component. 

There is one drawback though. From the point of view of the physical application, the boundary part of the diagram (consisting of the squids whose heads are not in the node relation with any other head - they form the boundary graph) describes either the initial and final state or, more generally,  the surface state. The remaining  part of the diagram consists of the pairs of the related nodes (internal edges) and describes the interaction. The two pieces of this information are entangled in the presented characterization. The reason is, that the diagrams, obtained with the algorithm above from a given squid set endowed with the node and link relations (the data $(i)-(iii)$) by implementing all the possible glueings (step $(iv)$), in general have different boundary graphs (the graphs share a squid set, but have different graph structures). Therefore we do not control in that way the boundary of the diagram, that is the initial-final/boundary Hilbert space. We  improve that characterization and the algorithm in the next section.

\section{Operator spin-network diagrams with fixed boundary}\label{sc:piany_o_ustalonym_brzegu}

In this section we  improve the construction and the algorithm introduced in  the previous section. The improved algorithm provides all the operator spin-network diagrams which have a same, {\it arbitrarily fixed  boundary graph}. 

\subsection{The idea and the trick}
In \cite{OSD} (see Sec.~6.2), for every graph $\Gamma$ we introduced the graph diagram corresponding to the {\it static spin-foam}, that is the spin-foam describing the trivial evolution of the graph.  Let us denote this diagram by ${\cal D}_\Gamma$ and refer to it as the {\it static graph diagram of} $\Gamma$. The boundary of ${\cal D}_\Gamma$ is the disjoint union $\Gamma\cup\bar{\Gamma}$ where $\bar{\Gamma}$ is a graph obtained by switching the orientation of each of the links of $\Gamma$ (\fref{2_10_trivial_diagram}).
\begin{figure}[htb!]
	\centering
	\subfloat[$\;$]{\includegraphics[width=0.22\textwidth]{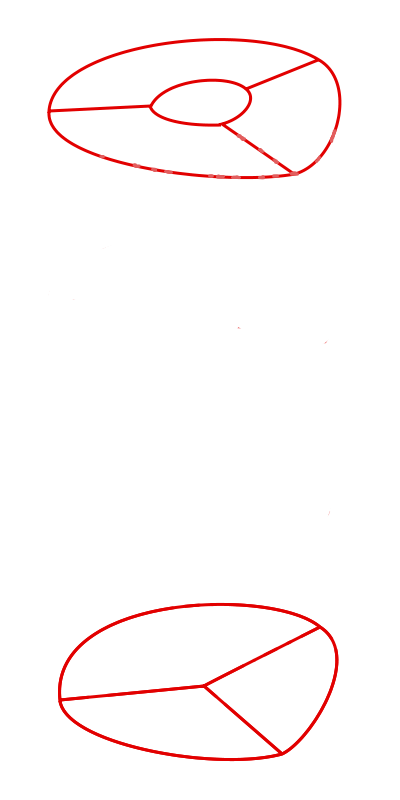}}
	\hspace{0.03\textwidth}
	\subfloat[$\;$]{\includegraphics[width=0.22\textwidth]{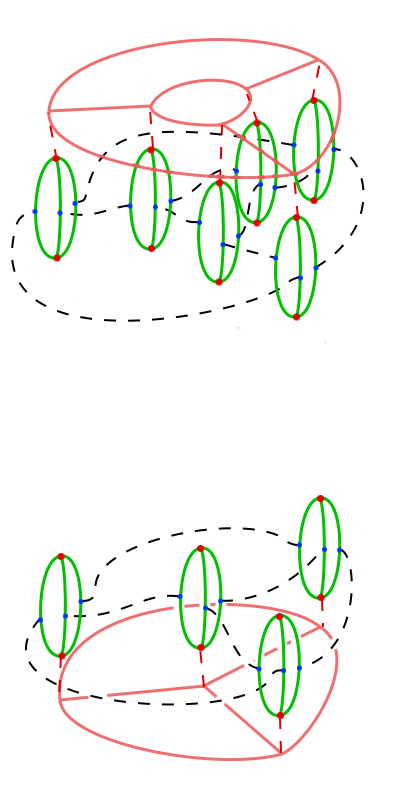}}
	\hspace{0.03\textwidth}
	\subfloat[$\;$]{\includegraphics[width=0.22\textwidth]{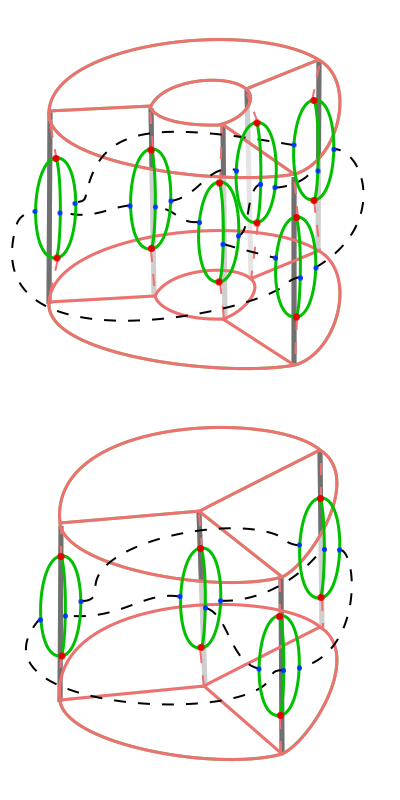}}
	\hspace{0.03\textwidth}
	\subfloat[$\;$]{\includegraphics[width=0.22\textwidth]{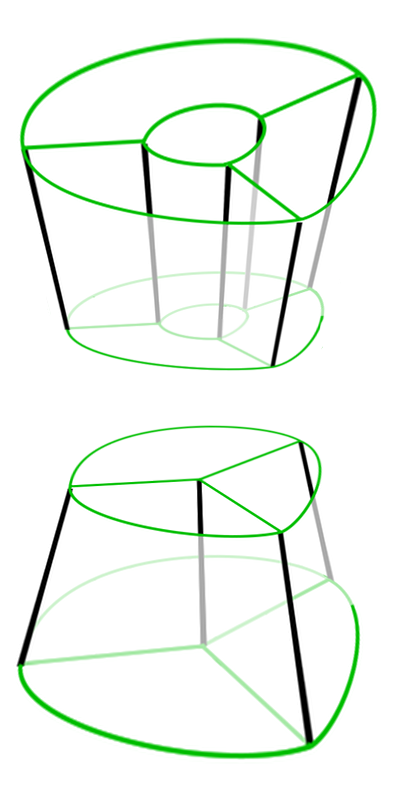}}
	\caption{A static diagram. $(a)$ A given graph $\Gamma$ consisting of two connected components. $(b)$ The corresponding static graph diagram ${\cal D}_{\Gamma}$ built of $\theta$ graphs and suitably defined node and link relations. $(c)$ The scheme of building a static  foam from the diagram. $(d)$ The boundary graph $\partial {\cal D}_{\Gamma}$ of the resulting foam is the disjoint union of $\Gamma$ and $\overline\Gamma$.}
	\label{fig:2_10_trivial_diagram}
\end{figure}

Given a coloring of the links of the graph $\Gamma$ by representations,  
we endow the static graph diagram ${\cal D}_\Gamma$ with the natural colorings
(see \cite{OSD}, Sec.~6.2): $(i)$
the coloring $\rho$  of the links of the graphs in ${\cal D}_\Gamma$ is the one induced by the coloring of the links of $\Gamma$; $(ii)$ the operator coloring $P$ colors with the identity operators; $(iii)$ finally, all the contractor coloring $A$ assigns to each $n$-theta graph of ${\cal D}_\Gamma$  the natural trace contractor $A^{\rm Tr}$. The result is  the {\it static OSN-diagram}. The corresponding operator is the identity in the Hilbert space given by the coloring of the links of $\Gamma$. The technical definition of  static diagram will be recalled in the next subsection.
       
The diagrams we will construct with the improved algorithm introduced below, contain the boundary $\Gamma$ together with its static diagram ${\cal D}_\Gamma$ as a subdiagram. Given $\Gamma$, our construction will provide all the diagrams of this type, that is all the diagrams, modulo the static sub-diagram. In terms of the spin-foam formalism, we will construct all the spin-foams  bounded by an arbitrarily fixed graph $\Gamma$. In each of the spin-foams, the neighborhood of $\Gamma$ is homeomorphic to the cylinder $\Gamma\times$ $[0,1]$.

\begin{figure}[bt!]
	\centering
	\includegraphics[width=0.37\textwidth]{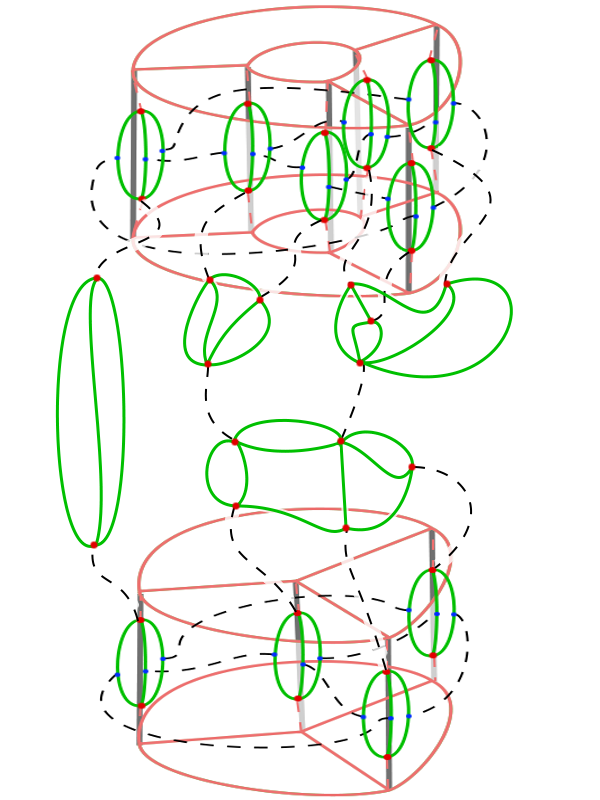}
	\caption{The static diagram ${\cal D}_{\Gamma}$ of \fref{2_10_trivial_diagram} glued to an interaction diagram ${\cal D}_{\rm int}$. The result is a diagram ${\cal D}_{\Gamma}\#{\cal D}_{\rm int}$ (link relations are omitted).}
	\label{fig:2_11_trivial_att}
\end{figure}
The key trick behind our construction is the following observation: given a graph $\Gamma$ of the squid set ${\cal S}_\Gamma$,  and an arbitrary graph diagram ${\cal D}_{\rm int}$ called {\it interaction diagram}, whose boundary graph $\partial{\cal D}_{\rm int}$ has the {\it same squid set}:
\begin{equation}
	{\cal S}_{\partial{\cal D}_{\rm int}}\ =\ {\cal S}_{\Gamma}
\end{equation}
we can combine the diagram ${\cal D}_{\rm int}$ with the static diagram  ${\cal D}_\Gamma$  into the new graph diagram ${\cal D}_{\Gamma}\# {\cal D}_{\rm int}$ such that the graph $\Gamma$ becomes its boundary. We achieve that by defining the graph of ${\cal D}_{\Gamma}\# {\cal D}_{\rm int}$ to be the disjoint union of the graph of ${\cal D}_\Gamma$ with the graph of ${\cal D}_{\rm int}$ and  extending the node and link relations of the component diagrams such that each squid of the boundary ${\partial\cal D}_{\rm int}$ is related to the corresponding squid  of $\bar{\Gamma}$ (a part of the boundary of ${\cal D}_\Gamma$) - see \fref{2_11_trivial_att}.

The identification of the squid sets ${\cal S}_{\Gamma}$ and $\partial{\cal D}$ 
is defined modulo  symmetries of ${\cal S}_{\Gamma}$ (exchanging identical squids). 
In the consequence, the graphs $\bar{\Gamma}$ and $\partial{\cal D}_{\rm int}$ may admit more than one way of relating their squid. In the algorithm below, the identification (a bijection) between the squid sets will be given, and the freedom will be in the glueing of the legs.

Below we implement this idea in detail, beginning with recalling the exact definition
of the static diagrams. 

\subsection{Static diagrams}
First, we recall the definition of the static diagram of a graph $\Gamma$. We use the squid set ${\cal S}_\Gamma$. For each squid $\lambda\in{\cal S}_\Gamma$ we introduce a theta-like graph $\tilde{\theta}_\lambda$ as follows (\fref{2_12_squid_to_theta}):
\begin{figure}[ht!]
	\centering
	\includegraphics[width=0.60\textwidth]{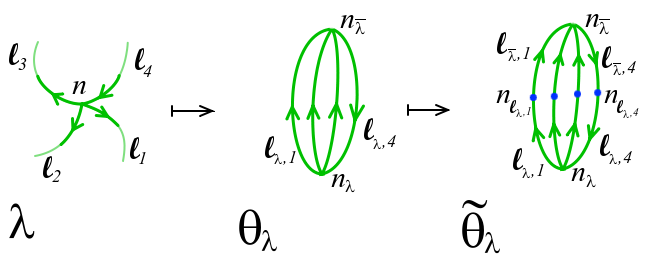}
	\caption{The procedure of creating a $\theta$-like graph from a squid $\lambda$.}
	\label{fig:2_12_squid_to_theta}
\end{figure}
\begin{itemize}
	\item $\bar{\lambda}$: we introduce a new squid $\bar{\lambda}$ obtained by flipping the orientation of each of the legs of $\lambda$. 
	\item $\theta_\lambda$: glue each  outgoing/incoming leg $\ell_{\lambda,i}$ of $\lambda$ with the corresponding incoming/outgoing leg ${\ell}_{\bar{\lambda},i}$ of $\bar{\lambda}$ to obtain a link ${\ell}_{\bar{\lambda},i}\circ \ell_{\lambda,i}$ /  $\ell_{\lambda,i}\circ {\ell_{\bar{\lambda},i}}$  connecting the head $n_\lambda$ of the squid $\lambda$ with the head ${n}_{\bar{\lambda}}$ of the squid $\bar{\lambda}$; the result is a closed graph $\theta_\lambda$.
\item $\tilde{\theta}_{\lambda}$: on each link $\bar{\ell}_{\lambda,i}\circ \ell_{\lambda,i}$ /  $\ell_{\lambda,i}\circ \bar{\ell}_{\lambda,i}$ we introduce an extra node $n_{\lambda,i}$; the resulting graph is denoted $\tilde{\theta}_\lambda$
\end{itemize}
The links of the graph $\tilde\theta_\lambda$ are just the legs $\ell_{\lambda,i}$, $\bar{\ell}_{\lambda,i}$, $i=1,2,...$ of the squids $\lambda$ and $\bar{\lambda}$ respectively.

The result of this procedure is a family of the graphs $\tilde{\theta}_\lambda$, one per each squid $\lambda\in {\cal S}_\Gamma$. The disjoint union $\coprod_\lambda  \tilde{\theta}_\lambda$ is the graph of the static graph diagram ${\cal D}_\Gamma$. The node and the link relations in $\coprod_\lambda  \tilde{\theta}_\lambda$ are induced by the the structure of the graph $\Gamma$ - see \fref{2_13_trivial}.
\begin{figure}[hbt!]
	\centering	\subfloat[\label{fig:2_13_trivial_1}]{\includegraphics[width=0.3\textwidth]{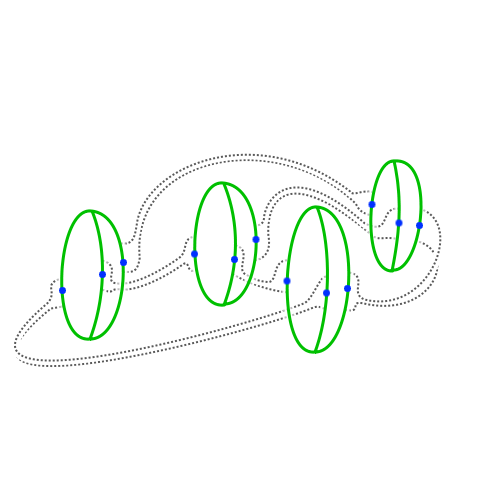}}
	\hspace{0.1\textwidth}
	\subfloat[\label{fig:2_13_trivial_2}]{\includegraphics[width=0.3\textwidth]{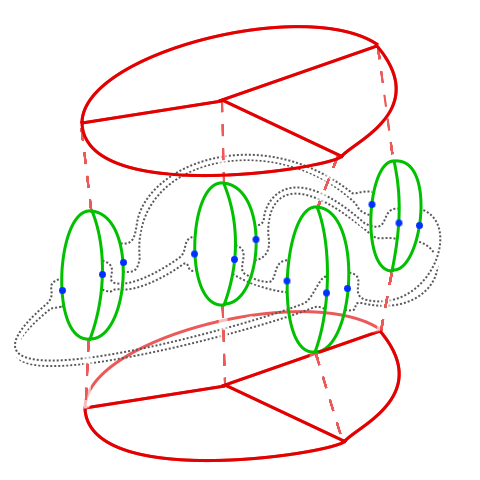}}
	\caption{ (a) A static diagram. (b)  Its boundary graph.}
	\label{fig:2_13_trivial}
\end{figure}

To define the node relation, notice that the set of the nodes of the graph $\coprod_\lambda  \tilde{\theta}_\lambda$ consists of the nodes of two types:
\begin{itemize} 
	 \item The first type is the heads of the squids $\lambda\in{\cal S}_\Gamma$ (denoted by $n_\lambda$) and the heads of the conjugate squids $\bar{\lambda}$ (denoted by $n_{\bar{\lambda}},...$) - those nodes are left unrelated by the node relation, that is they become the boundary nodes.

	\item The ends of the  legs of the squids $\lambda\in{\cal S}_\Gamma$ (glued with the legs of the conjugate squids $\bar{\lambda}$) denoted by $n_{\ell_{\lambda,i}}$.
 Whenever $\ell_{\lambda,i}\circ\ell'_{\lambda',i'}$ is a link of $\Gamma$, then
the nodes $n_{\ell_{\lambda,i}}$ and $n_{\ell'_{\lambda',i'}}$ are in the node relation.
\end{itemize}
The link relation is defined as follows:
\begin{itemize}
	\item  For every pair of  the  nodes $n_{\ell_{\lambda,i}}$ and  $n_{\ell'_{\lambda',i'}}$ related above by the node relation, each node is two-valent. The link relation is defined to pair the links $\ell_{\lambda,i}$ and $\ell'_{\lambda',i'}$, as well as the links $\ell_{\bar{\lambda},i}$ and $\ell'_{\bar{\lambda}',i'}$ (see the previous item).
\end{itemize}

Given a static graph diagram ${\cal D}_{\Gamma}$ the natural coloring consists of: irreducible representations freely assigned to the links, the operators $P_{n_\lambda}={\rm id}:{\cal H}_{n_\lambda}\rightarrow{\cal H}_{n_\lambda}$ for every head $n_\lambda$ and every squid $\lambda$, and $P_{n_{\ell_{\lambda,i}}n_{\ell'_{\lambda',i'}}}$ equal to the natural isomorphisms ${\rm id}:{\cal H}_{n_{\ell_{\lambda,i}}}\rightarrow{\cal H}_{n_{\ell'_{\lambda',i'}}}$. Finally, the contractor assigned to each component graph $\tilde{\theta}_\lambda$ is the natural $A^{\rm Tr}_{\tilde{\theta}}$.

\subsection{The improved algorithm}\label{sc:algorithm}
We present now our algorithm for the construction of the operator spin-network  diagrams whose unoriented boundary is an arbitrarily fixed unoriented graph $|\Gamma|$. 
\begin{enumerate} 
	\item $\Gamma$: choose an orientation of each link of $|\Gamma|$, the result is an oriented graph $\Gamma$ (\fref{2_14_g2g}).
	\begin{figure}[ht!]
		\centering
		\includegraphics[width=0.5\textwidth]{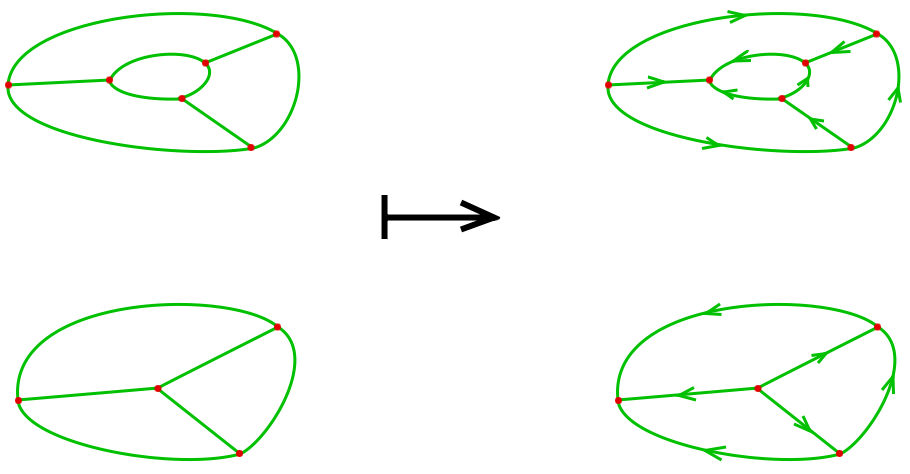}
		\caption{$|\Gamma|$ on the LHS and $\Gamma$ on the RHS.}
		\label{fig:2_14_g2g}
	\end{figure}
	
	\item ${\cal S}_{\rm int}$: to the squid set ${\cal S}_{\Gamma}$ of the graph $\Gamma$ add $N$ pairs of squids $\lambda_1,\lambda'_1,...,\lambda_{N},\lambda'_{N}$, such that for each pair the number of incoming/outgoing legs of one squid equals the number of outgoing/incoming legs of the other one (\fref{2_15_squid_set_extendet}). Denote the resulting squid set ${\cal S}_{\rm int}$. Introduce a node relation by relating the head of $\lambda_i$ with  the head of $\lambda'_i$, $i=1,...,N$. 
	\begin{figure}[ht!]
		\centering
		\subfloat[\label{fig:2_15_squid_set_1}]{\includegraphics[width=0.16\textwidth]
				{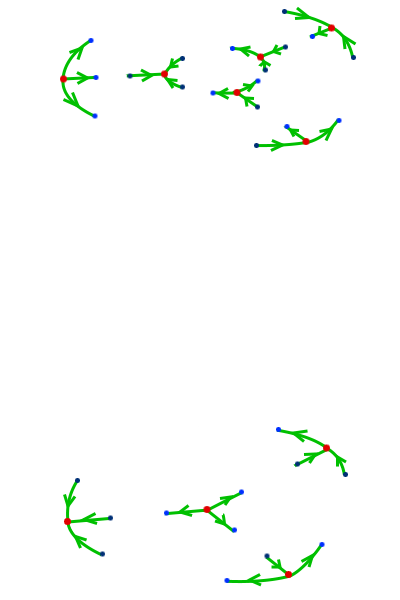}}
		\hspace{0.2\textwidth}
		\subfloat[\label{fig:2_15_squid_set_2}]{\includegraphics[width=0.16\textwidth]
						{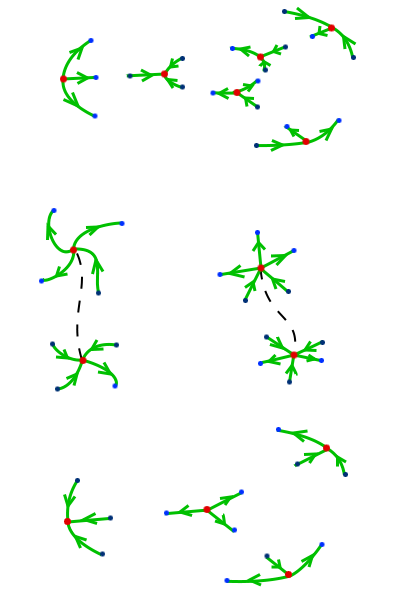}}
		\caption{(a) ${\cal S}_\Gamma$, and ${\cal S}_{\rm int}$ in the case of $N=0$ (b) ${\cal S}_{\rm int}$ obtained by adding N=2  pairs of squids paired by a node relation.}
		\label{fig:2_15_squid_set_extendet}
	\end{figure}

	\item ${\cal D}_{\rm int}$: To the squid set ${\cal S}_{\rm int}$ with the chosen set of pairs of the squids 	$\lambda_1,\lambda'_1,...,\lambda_{N},\lambda'_{N}$ and the node relation apply the steps $(iii)$ and $(iv)$ of the algorithm of \sref{algorithm1}. Denote the resulting graph diagram ${\cal D}_{\rm int}$ (\fref{2_16_graph_diagram}).
	\begin{figure}[ht!]
		\centering
		\subfloat[\label{fig:2_16_graph_diagram_1}]
				{\includegraphics[width=0.16\textwidth]{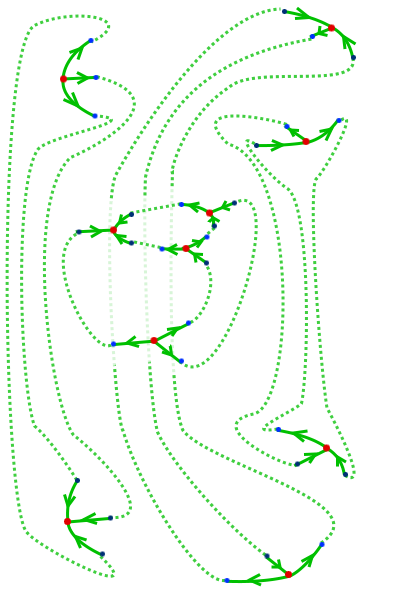}}
		\hspace{0.2\textwidth}
		\subfloat[\label{fig:2_16_graph_diagram_2}]
				{\includegraphics[width=0.16\textwidth]{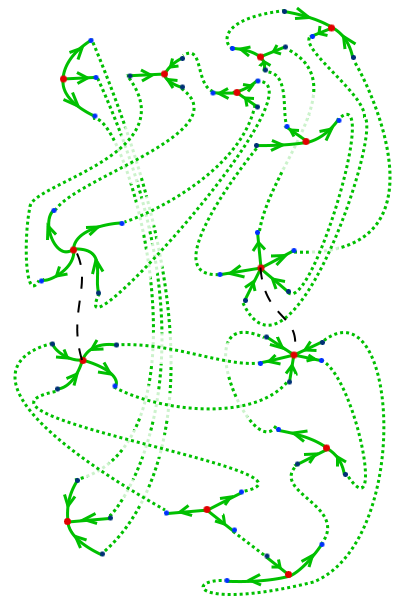}}
		\caption{Step $(iii)$ of the improved algorithm: glueing of a graph diagram ${\cal D}_{\rm int}$ (one of several possible) from the squid set ${\cal S}_{\rm int}$ presented
		at \fref{2_15_squid_set_1}  and, respectively, \fref{2_15_squid_set_2}. The dotted lines mark the glueing the legs of the squids.}
		\label{fig:2_16_graph_diagram}
	\end{figure}
 
	\item  ${\cal D}_{\Gamma}\#{\cal D}_{\rm int}$: Use the static graph diagram ${\cal D}_\Gamma$ of the graph $\Gamma$ and construct the union of the diagrams as it was explained above (\fref{2_17_glue_trivial}).
	\begin{figure}[htb!]
		\centering
		\subfloat[$\;$]{\includegraphics[width=0.17\textwidth]{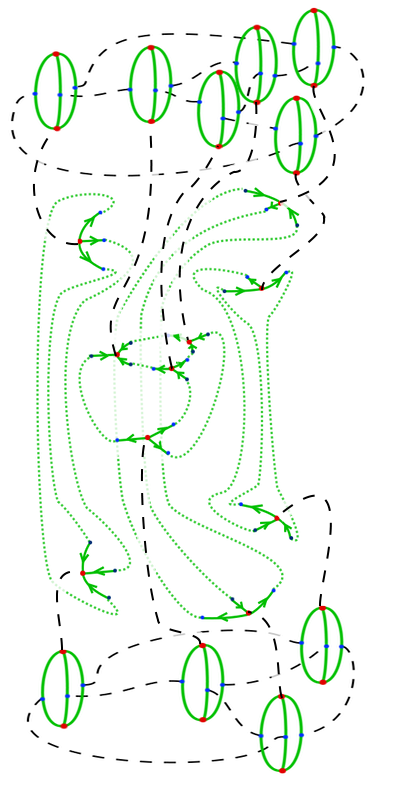}\label{fig:2_17_glue_trivial_1}}
		\hspace{0.2\textwidth}
		\subfloat[$\;$]{\includegraphics[width=0.17\textwidth]{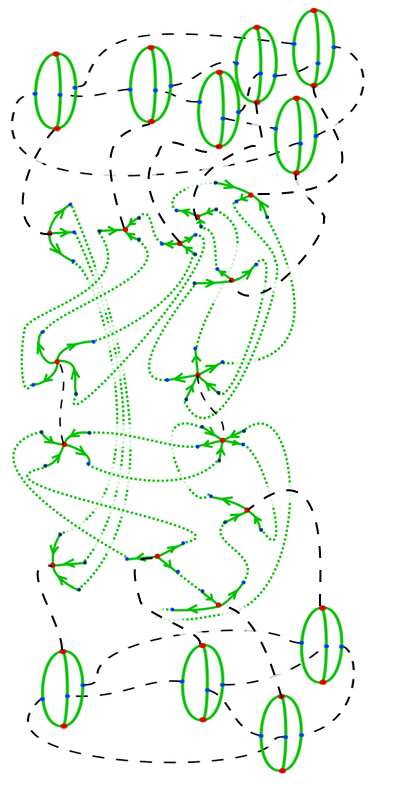}}
		\caption{Step $(iv)$ of the improved algorithm: the graph diagram ${\cal D}_{\rm int}$ of		\fref{2_16_graph_diagram_1} and, respectively, \fref{2_16_graph_diagram_2} combined with the static diagram ${\cal D}_{\Gamma}$ of the graph $\Gamma$ of \fref{2_14_g2g}, into the final graph diagram ${\cal D}_{\Gamma}\#{\cal D}_{\rm int}$.}   
		\label{fig:2_17_glue_trivial}
	\end{figure}

		\item Coloring: Define arbitrary coloring of the diagram ${\cal D}_{\Gamma}\#{\cal D}_{\rm int}$ which turns it into operator an spin-network diagram (\fref{2_18_coloring}).
	\begin{figure}[h!]
		\centering
		\subfloat[\label{fig:2_18_coloring_1}]{\includegraphics[width=0.23\textwidth]{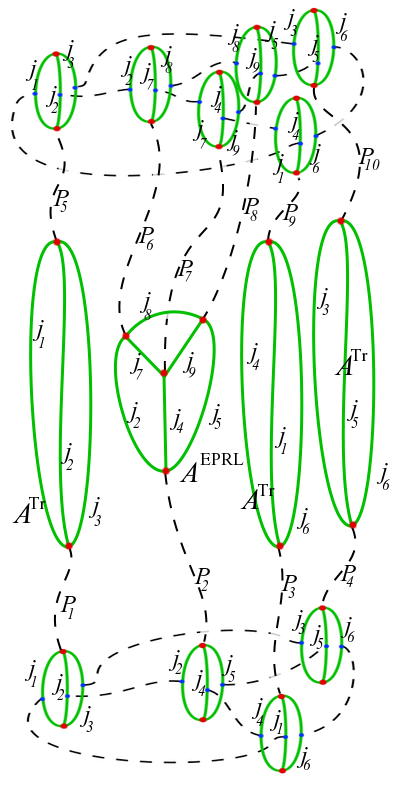}}
		\hspace{0.1\textwidth}
		\subfloat[\label{fig:2_18_coloring_2}]{\includegraphics[width=0.23\textwidth]{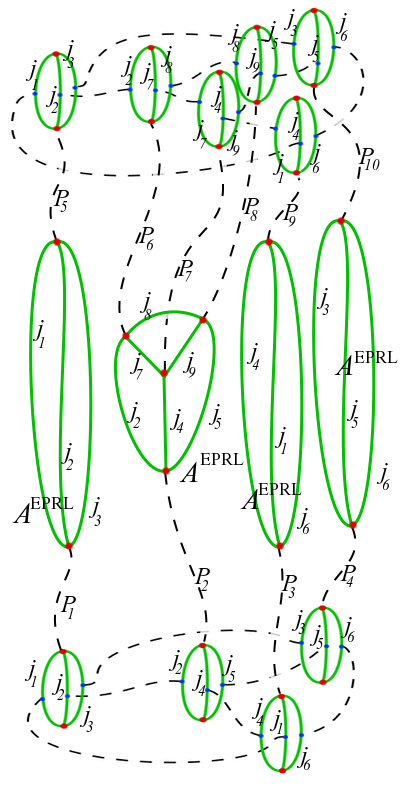}}
		\caption{The graph diagram of \fref{2_17_glue_trivial_1} colored in two possible ways. The coloring of the boundary trivial diagram by operators and contractors is omitted for transparency of the figure (each pair of nodes in this part of diagram is colored by appropriate $\id_j$, each graph is colored by $A_\theta^{\Tr}$). The coloring of (a) represents diagram with trivial evolution on the vertical edges. The coloring of (b) represents diagram with EPRL norm calculated for each vertical edge.}
		\label{fig:2_18_coloring}
	\end{figure}

		\item Consider all possible: orientations of $|\Gamma|$, $N$-tuples of pairs of squids added to ${\cal S}_{\Gamma}$, ways of connecting the legs of ${\cal S}_{\rm int}$, link relations for each $\lambda_i,\lambda'_i$, colorings in $(v)$.

\end{enumerate}
Notice, that it would be insufficient to fix one orientation of the boundary. A priori all the orientations have to be taken into account. On the other hand, in general, the algorithm will possibly give OSN-diagrams related by the reorientation (see the previous section). In specific cases, that redundancy should be reduced. The presented construction allows to control  the level of complexity of resulting diagram by the level of complexity of  the diagram ${\cal D}_{\rm int}$. The complexity can be measured by the number of pairs of the nodes related by the node relation, that is the number of internal edges. The simplest case is zero internal edges, that is the squid set 
\be{\cal S}_{\rm int}\ =\ {\cal S}_\Gamma\ee
In that case all the  graph used to define  graph diagram ${\cal D}_{\rm int}$ becomes the boundary graph $\partial{\cal D}_{\rm int}$.  

A general example of the interaction graph diagram ${\cal D}_{\rm int}$ is given by a graph $\tilde{\Gamma}_{\rm int}$ and a node relation ${\cal R}_{\rm int}^{\rm node}$ consisting of exactly $N$ pairs of nodes. Increasing the number $N$ we increase the complexity of the diagram ${\cal D}_{\Gamma}\# {\cal D}_{\rm int}$.

In the previous subsection we have recalled the structure of colorings that turn graph diagrams into operator spin-network diagrams. In the case of the graph diagrams constructed in this subsection, without lack of generality, it is sufficient to consider colorings of the diagrams ${\cal D}_\Gamma\#{\cal D}_{\rm int}$  provided by our  algorithm, which reduced to the static graph diagram ${\cal D}_\Gamma$ provides the static operator spin-network diagram defined above.       A way to control the freedom in the colorings by representations is to fix a coloring of the links of the boundary graph $\Gamma$.

In the next section we apply the algorithm to construct all the operator spin-network diagrams of the Rovelli-Vidotto dipole cosmology.

\StopkaPliku

 \ifx \NaglowekPliku \undefined
  \def \NaglowekPliku {
\documentclass[notitlepage]{iopart}
\begin{document}
}
\fi
\ifx \StopkaPliku \undefined
  \def \StopkaPliku {\end{document}}
\fi
\NaglowekPliku

\section{Diagrams with boundary given by dipole cosmology graph}\label{sc:dipole_cosmology}
In this section we apply the algorithm presented above to a specific example of a boundary graph, namely the one given by the Dipole Cosmology model of Bianchi, Rovelli and Vidotto \cite{SF_cosmology}.

The Dipole Cosmology model is an attempt to test the behaviour of the spin-foam transition amplitudes in the limit of the homogeneous and isotropic boundary states. The boundary state of this model is supported on so called \emph{dipole graph} (\fref{thetas}) which consists of two disjoint components, $4$-valent theta graphs  representing the initial and, respectively, final geometry. The transition amplitude is calculated under several approximations. One of the approximations is the vertex expansion - i.e. at the first order one considers contribution of the spin-foams with four internal edges and one interaction vertex only.

\subsection{The improved algorithm in the Dipole Cosmology case}
We apply now the algorithm of the previous section to construct all the operator spin-network diagrams whose boundary is fixed (modulo an orientation) to be the graph $|\Gamma|$ which consists of two disjoint $4$-valent theta graphs $|\Gamma_{4\theta}|$ (\fref{thetas}), 
\be
	|\Gamma| \ =\ |\Gamma_{4\theta}| \cup |\Gamma_{4\theta}|,
\ee
and which correspond to  1-vertex spin-foams.
\begin{figure}[ht!]
	\centering
	\includegraphics[width=0.15\textwidth]{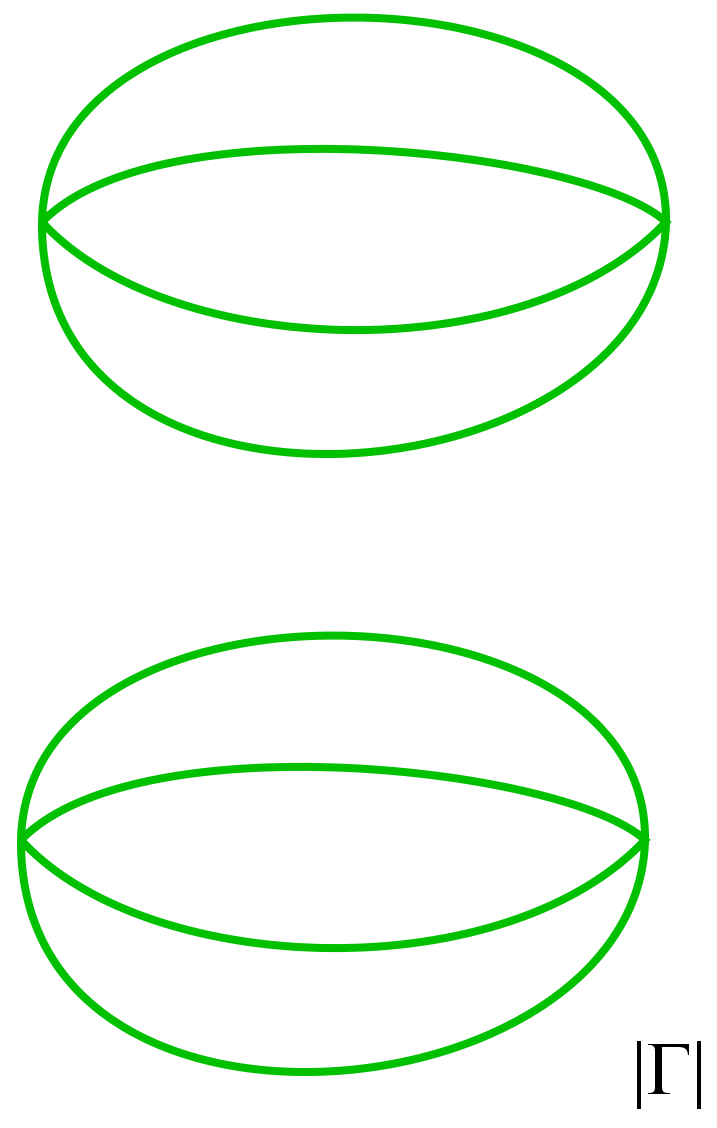}
	\caption{The dipole boundary graph $|\Gamma|=|\Gamma_{4\theta}| \cup |\Gamma_{4\theta}|$.}
	\label{fig:thetas}
\end{figure}

In terms of the improved algorithm of the previous section this assumption means that the the interaction diagram ${\cal D}_{\rm int}$ consists of one graph $\Gamma_{\rm int}$ whereas the node and link relations are trivial. That is its squid set equals the squid set of the boundary (the initial plus the final) graph. Specifically, for this Dipole Cosmology example and with the 1-vertex assumption the improved algorithm from \sref{algorithm} reads:
 
\begin{enumerate} 
	\item\label{orientacja} $\Gamma$: choose an orientation of each link of each of the two graphs $|\Gamma_{4\theta}|$ (\fref{thetas_oriented}).
\begin{figure}[ht!]
	{\centering
	\begin{tabular}{>{\centering}p{0.3\textwidth}>{\centering}p{0.3\textwidth}>{\centering}p{0.3\textwidth}}
			\subfloat[$\;$\label{fig:thetas_oriented}]{\includegraphics[width=0.19\textwidth]{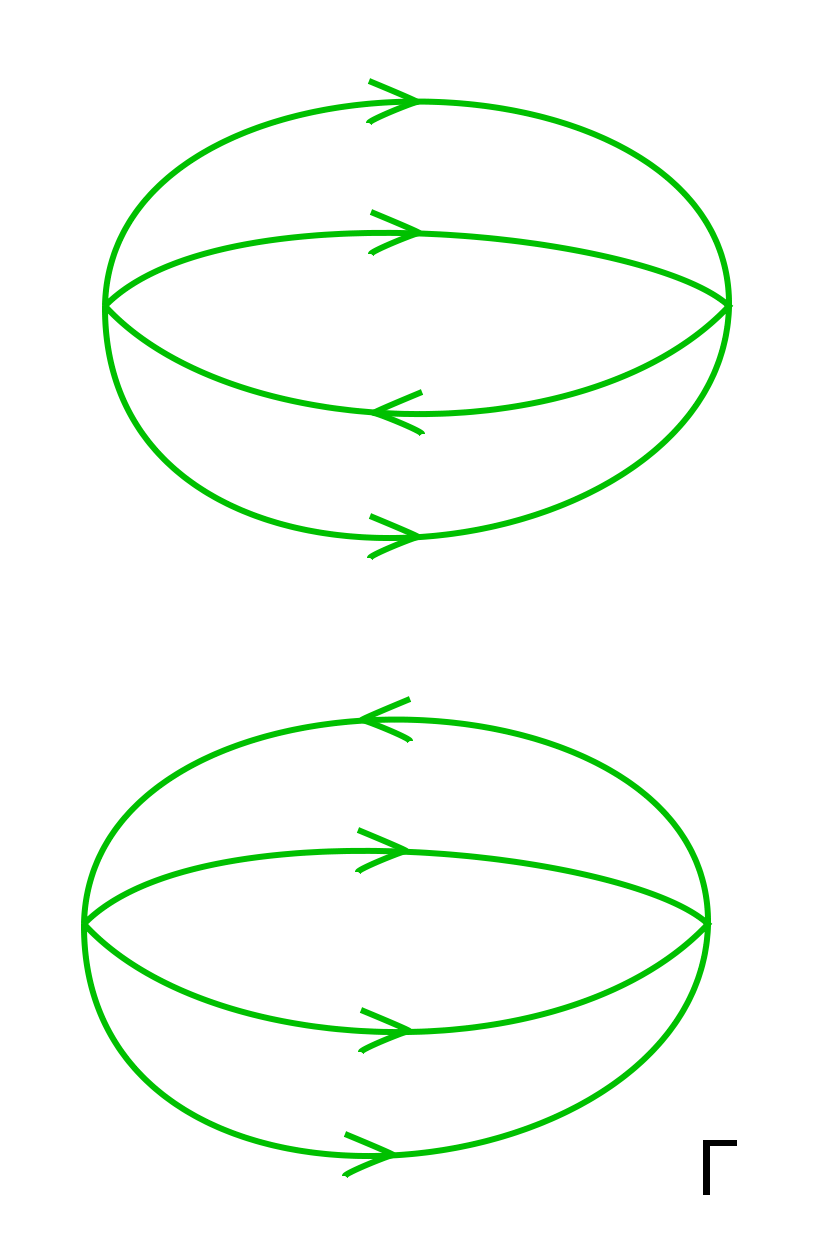}}                                                                        
&
			\subfloat[$\;$\label{fig:thetas_squid_set}]{\includegraphics[width=0.19\textwidth]{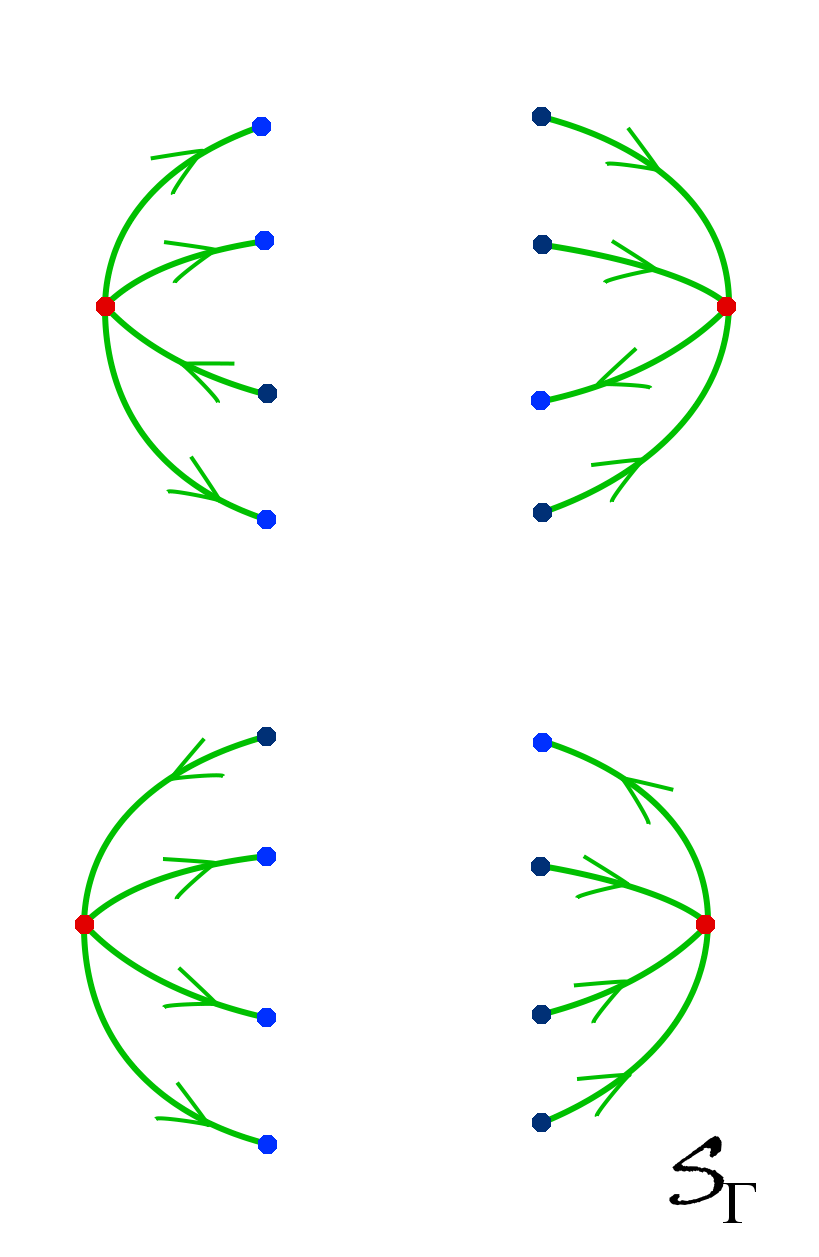}}
&
			\subfloat[$\;$\label{fig:interaction_graph}]{\includegraphics[width=0.19\textwidth]{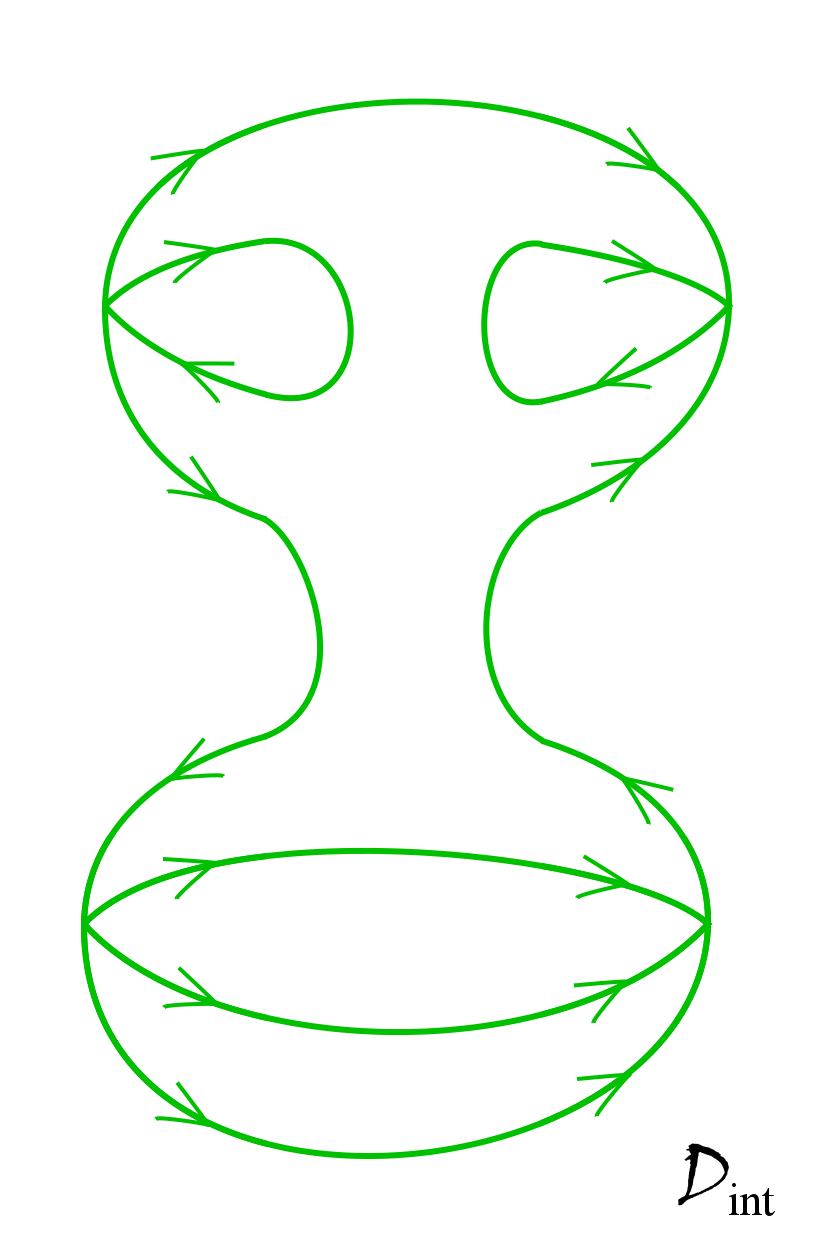}}
	\end{tabular}
	}
	\caption{Construction of the graph diagram. (a) Step $(i)$ -- choose an orientation of each link. (b) Step $(ii)$ -- construct the squid set ${\cal S}_{\Gamma}$. (c) Step $(iii)$ -- construct an interaction graph ${\rm D}_{\rm int}=\Gamma_{\rm int}$; an example is depicted.}
	\label{fig:construction_1}
\end{figure}

	\item ${\cal S}_{\rm int}={\cal S}_{\Gamma}$: for the interaction squid set take  the squid set ${\cal S}_{\Gamma}$ of the graph $\Gamma$; this means that in point $(ii)$ of the general algorithm presented in \sref{algorithm} we set $N=0$ -- in other words, we consider the first order of the vertex and edge expansion; the interaction squid set consists of four $4$-valent squids; two of them are oriented freely - their orientation  determines the orientation of the remaining two squids and defines the orientation of $\Gamma$ (\fref{thetas_squid_set}).

\item $\Gamma_{\rm int}$: glue each incoming/outgoing leg of each squid of ${\cal S}_{\rm int}$ with an outgoing/incoming leg of another (or the same) squid  of ${\cal S}_{\rm int}$. In the next subsection we construct and list all the possible (unoriented) interaction graphs (they are depicted later, on \fref{interaction_graphs}). $\Gamma_{\rm int}$ is obtained by assigning orientation to each link of one such graph (\fref{interaction_graph}). Together with the trivial node and link relations,
$\Gamma_{\rm int}$ defines an interaction graph ${\cal D}_{\rm int}$.

	\item  ${\cal D}_{\Gamma}\#{\cal D}_{\rm int}$: Use the static graph diagram ${\cal D}_\Gamma$ of the graph $\Gamma$ (\fref{static_graph_diagram}) and construct the union of the diagrams as it was explained above (\fref{thetas_graph_diagram}).
\begin{figure}[ht!]
	{\centering
	\begin{tabular}{>{\centering}p{0.5\textwidth}>{\centering}p{0.5\textwidth}}
			\subfloat[$\;$]{\label{fig:static_graph_diagram}\includegraphics[width=0.4\textwidth]{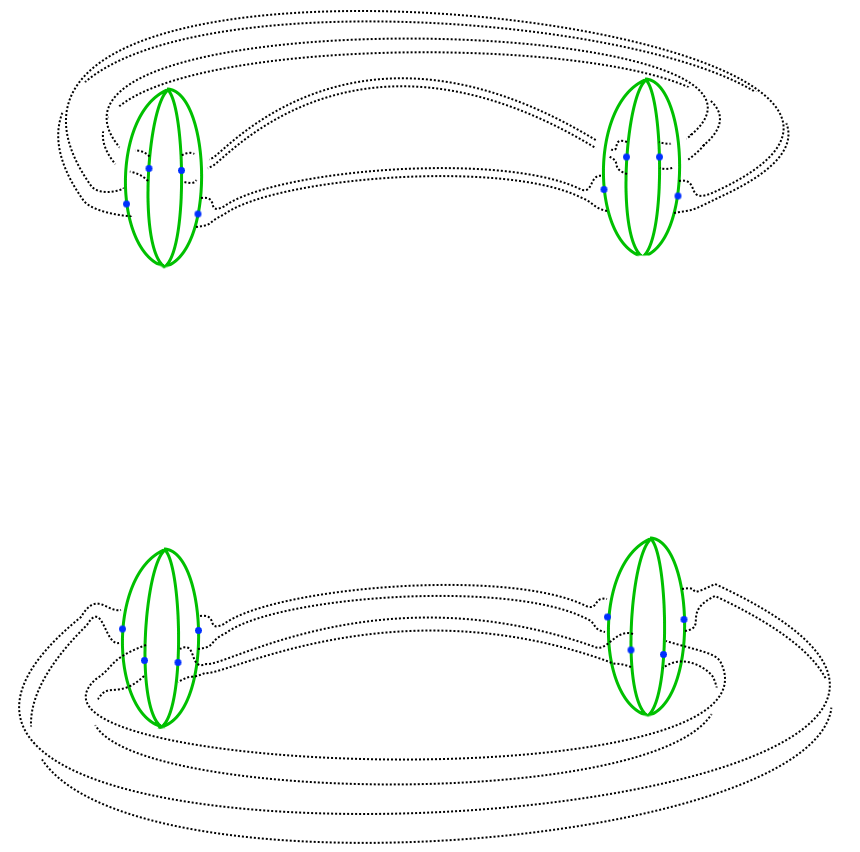}}
&
			\subfloat[$\;$]{\label{fig:thetas_graph_diagram}\includegraphics[width=0.4\textwidth]{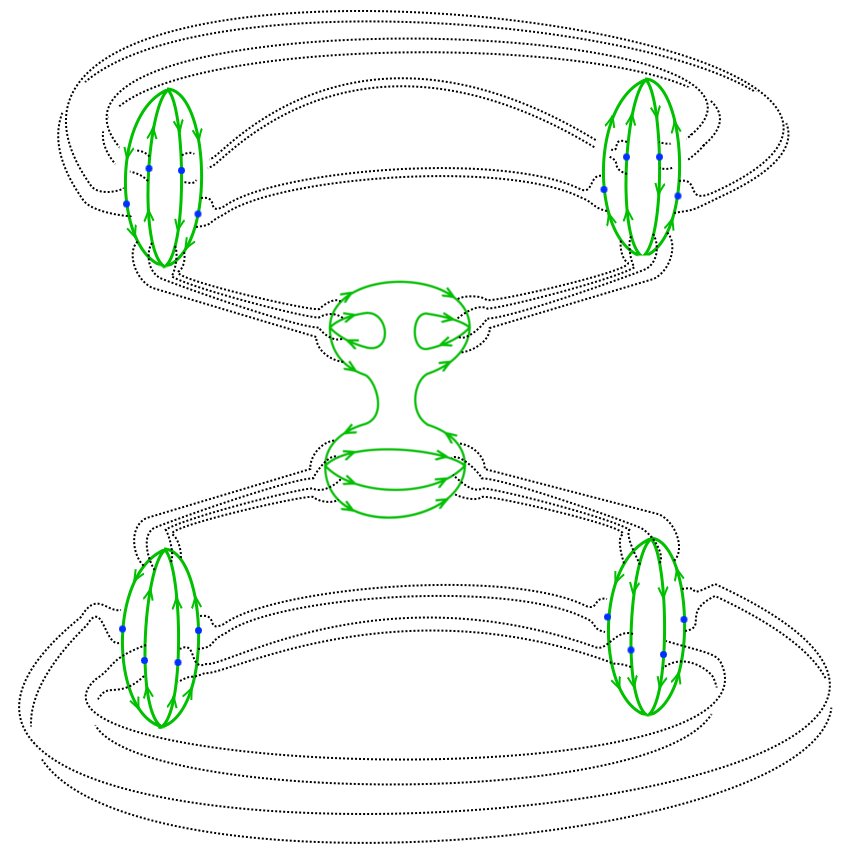}}
	\end{tabular}
	}
	\caption{Construction of the graph diagram. Step $(iv)$ -- the static graph diagram $\T D_\Gamma$ ((a) - the dotted lines denote the link relations) is attached to the diagram $\T D_{\rm int}$ (b), and the final diagram $\T D_\Gamma\#\T D_{\rm int}$ is obtained.}
	\label{fig:construction_2}
\end{figure}

		\item Coloring: Define arbitrary coloring of the diagram ${\cal D}_{\Gamma}\#{\cal D}_{\rm int}$ which turns it into an operator spin-network diagram.

		\item Consider all the possible: orientations of the two independent squids of ${\cal S}_{\Gamma}$, ways of connecting the legs of the four squids, all possible node relations between the nodes of the interaction graph and corresponding nodes of static diagram, all possible link relations between the links of the interaction graph and the corresponding links of the static diagram.
\end{enumerate}

\subsection{All the possible interaction graphs}
In this subsection we construct all the possible interaction graphs $\Gamma_{\rm int}$.  We obtain each interaction graph $\Gamma_{\rm int}$ by assigning an orientation to each link of an (unoriented) graph $|\Gamma_{\rm int}|$ defined by the following two properties:
\begin{itemize}
	 \item each graph $|\Gamma_{\rm int}|$ has exactly \textbf{4 nodes},
	 \item each node of $|\Gamma_{\rm int}|$ is precisely \textbf{four-valent}.
\end{itemize}
 We find below all possible graphs $|\Gamma_{\rm int}|$. We depicted the resulting graphs on \fref{interaction_graphs}. In order to obtain an interaction graph $\Gamma_{\rm int}$, we assign to each link of a graph $|\Gamma_{\rm int}|$ an orientation consistent with the orientation of the boundary (and the squid set). Given a graph $|\Gamma_{\rm int}|$ and (oriented) boundary graph $\Gamma$, such a choice of compatible orientation may be impossible. For example, take graph 1 from \fref{interaction_graphs} as a graph $|\Gamma_{\rm int}|$.  It is not possible to choose an orientation of links of this graph compatible with orientation of the boundary graph from \fref{thetas_oriented}. This is because the boundary graph \fref{thetas_oriented} has a node with three outgoing and one incoming link and such a structure of a node is not possible for graph 1 from \fref{interaction_graphs} (since each link of this graph forms a loop, the number of incoming links and outgoing links needs to be equal at every node). Note, that there is a distinguished graph $|\Gamma_{\rm int}|$ -- the graph 20 from \fref{interaction_graphs} used in \cite{SF_cosmology}. This graph may be oriented in a way compatible with any boundary graph $\Gamma$. The natural question which arises is whether for every graph from \fref{interaction_graphs} there is a boundary graph such that orientation of $|\Gamma_{\rm int}|$ may be chosen to be compatible with this boundary graph. The answer is affirmative. It may be shown that orientation of links of each graph $|\Gamma_{\rm int}|$ may be chosen to be compatible with a boundary graph oriented such that at every node a number of incoming links equals to a number of outgoing links.

We now present in details the construction of unoriented graphs possessing exactly 4 nodes, all of which are four-valent, i.e. all possible graphs $|\Gamma_{\rm int}|$. It is well known that each unoriented graph may be encoded in adjacency matrix. It is a symmetric matrix $A\in {\rm Sym}(n)$ with the number of columns/rows $n$ equal to the number of the vertices of this graph. The entries $A_{ij}$ are equal to the numbers of links connecting node $i$ with node $j$, with a specification that links forming closed loops (corresponding to diagonal entries) are counted twice. An example of such matrix and the corresponding graph is given on \fref{adjacency_matrix}.
\begin{figure}[ht!]
	\centering
	\includegraphics[width=0.80\textwidth]{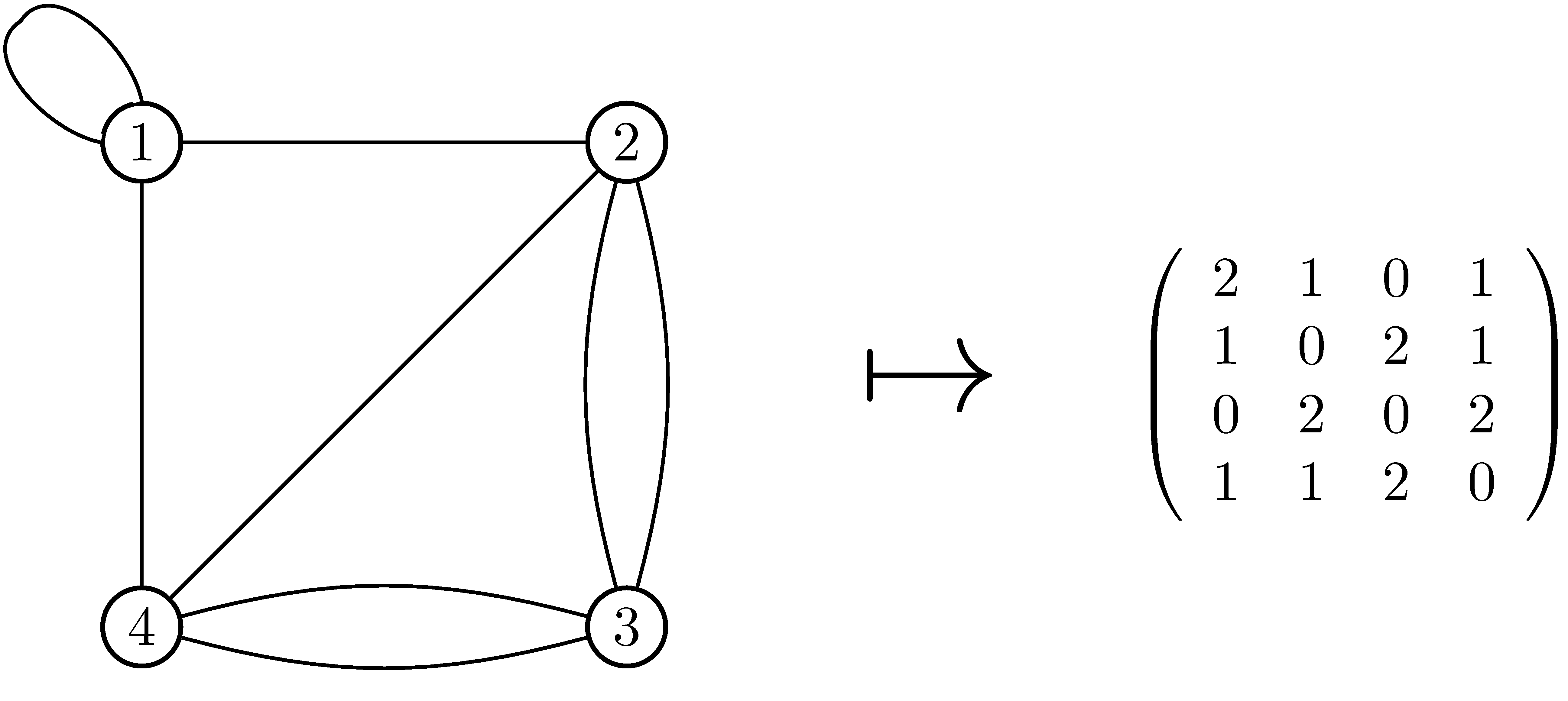}
	\caption{A graph and the corresponding adjacency matrix.}
	\label{fig:adjacency_matrix}
\end{figure}

However, given a graph there are many corresponding matrices, because for each permutation $\sigma\in {\rm S}_n$ the matrices
\be
 (\sigma \circ A)_{ij}:= A_{\sigma(i)\sigma(j)}
\ee
and $A_{ij}$ define the same graph. There is a natural bijective correspondence between graphs with $n$ vertices and orbits, elements of ${\rm Sym}(n)/S_n$.

In our case graphs have four nodes. We are therefore interested in $4\times 4$ matrices. The condition that each node is 4-valent corresponds to an assumption that the sum of numbers in each row/column is equal 4:
\begin{equation}\label{eq:czterowalencja}
	\forall_{i}\quad \sum_{j=1}^4 A_{ij}=4.
\end{equation}
The set of the possible interaction graphs $\mathcal{G}_{\rm int}$ is therefore characterised by the moduli space:
\be
	\left\{A\in {\rm Sym}(4): \forall_{i}\ \sum_{j=1}^4 A_{ij}=4\right\}/S_4.
\ee

First we introduce a parametrisation of the space of symmetric matrices satisfying \reef{czterowalencja} and then we find the moduli space using Wolfram's Mathematica 8.0.

To define our parametrisation in a transparent way, we introduce a triple $({\rm K}_4,d,m)$ (see \fref{graphs_notation}):
\begin{figure}[ht!]
	\centering
	\includegraphics[width=0.80\textwidth]{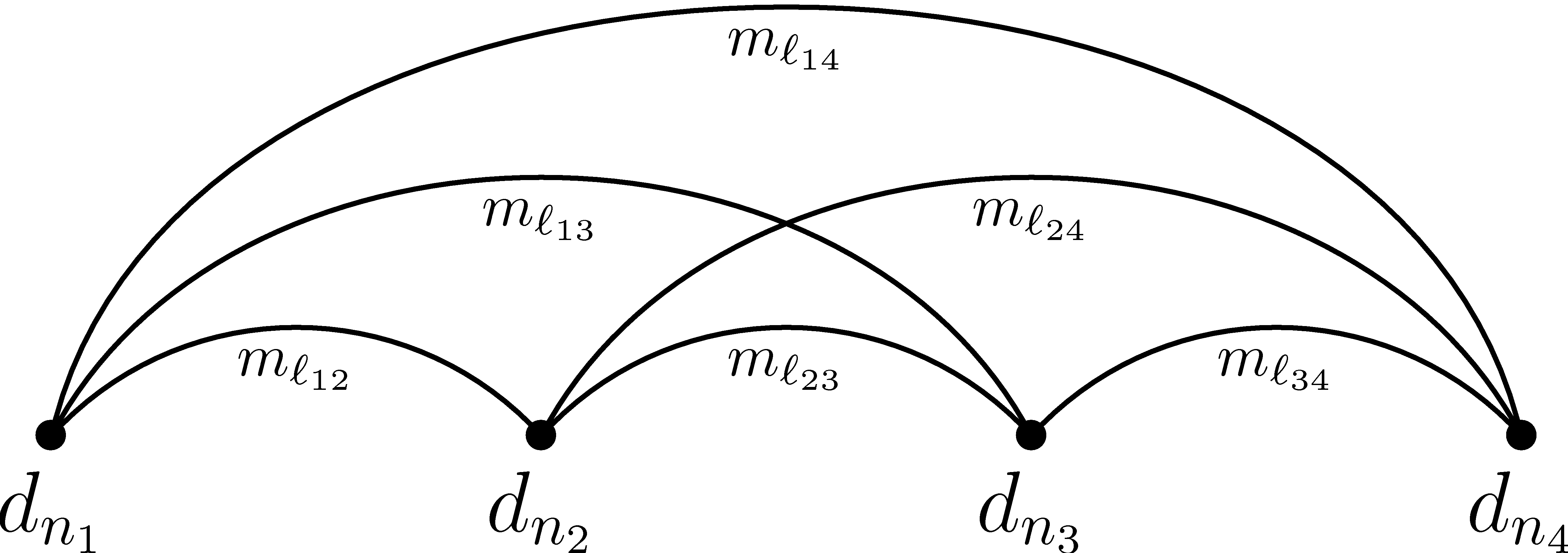}
	\caption{A graphical representation of $({\rm K}_4,d,m).$}
	\label{fig:graphs_notation}
\end{figure}
\begin{itemize}
	\item the complete graph ${\rm K}_4$ on four nodes -- the skeleton of a 4-simplex (we denote by ${\rm K}_4^{(0)}$ the set of its nodes and by ${\rm K}_4^{(1)}$ the set of its links);
	\item labeling of its nodes 
	\be
		 d:{\rm K}_4^{(0)}\ni n\mapsto d_n\in\{0,2,4\}
	\ee
	such that for all four nodes $n_1,\,n_2,\,n_3,\,n_4$ of the graph ${\rm K}_4$ the numbers $d_{n_1},\,d_{n_2},\,d_{n_3},\,d_{n_4}$ satisfy the generalized triangle inequalities:
	\be\label{eq:triangle_inequality}
		\forall_i\ d_{n_i}\leq \sum_{i\not=j} d_{n_j}.
	\ee
	\item labeling of its links
	\be
		m:{\rm K}_4^{(1)}\ni\ell\mapsto m_\ell\in \{0,1,2,3,4\}
	\ee
	such that
	\be\label{eq:summing}
		\forall_{n\in {\rm K}_4^{(0)}}\ \sum_{\{\ell\in {\rm K}_4^{(1)}:\ell\cap n \not= \emptyset\}} m_\ell = d_n
	\ee
\end{itemize}

The condition that $d_{n},\ n\in{\rm K}_4^{(0)}$ satisfy the generalized triangle inequalities \reef{triangle_inequality} ensures the existence of at least one labeling $m$\footnote{This well known fact is used for example in the representation theory of SU(2) to construct  of the invariants of the tensor product $\Hil_{{d_{n_1}}/{2}}\otimes\Hil_{{d_{n_2}}/{2}}\otimes\Hil_{{d_{n_3}}/{2}}\otimes\Hil_{{d_{n_4}}/{2}}$, where dim$\,\Hil_j=2j+1$, and $d_{n_1}+...+d_{n_4}\in 2\mathbb{N}$ by the construction.} 

To each triple $({\rm K}_4,d,m)$ corresponds a (multi)graph $\Gamma_{({\rm K}_4,d,m)}$, defined in the following way (see also an example at \fref{graphs_correspondence}):
\begin{figure}[hbt!]
	\centering
	\includegraphics[width=0.95\textwidth]{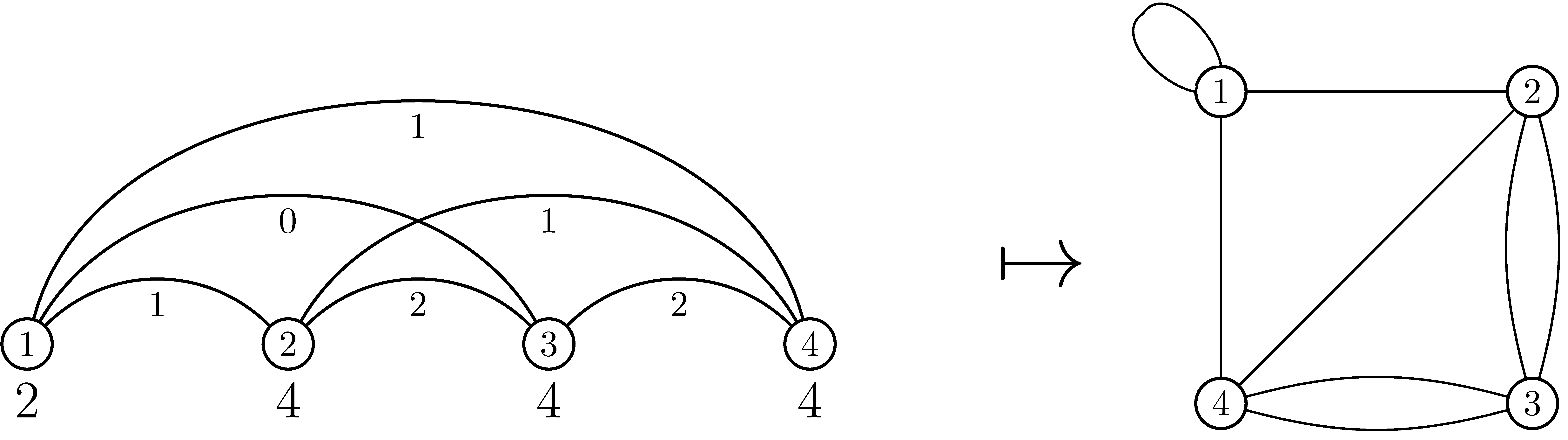}
	\caption{An example of the correspondence between $({\rm K}_4,d,m)$ and $\Gamma_{({\rm K}_4,d,m)}$. The numbering of the nodes is redundant here. However we add it to make the exposition clearer.}
	\label{fig:graphs_correspondence}
\end{figure}

\begin{itemize}
	 \item it has the same set of nodes $\Gamma_{({\rm K}_4,d,m)}^{(0)}={\rm K}_4^{(0)};$
	 \item for each pair $(n,n')$ of different nodes there are exactly $m_{\ell}$ links of $\Gamma_{({\rm K}_4,d,m)}$ connecting the nodes $n$ and $n'$, where $\ell$ is the link of $K_4$ connecting $n$ with $n'$ .
	 \item at each node $n$ there are precisely $(4-d_{n})/2$ links each of which makes a loop connecting $n$ with itself (\fref{node_notation});
\end{itemize}
\begin{figure}[ht!]
	\centering
	\includegraphics[width=0.9\textwidth]{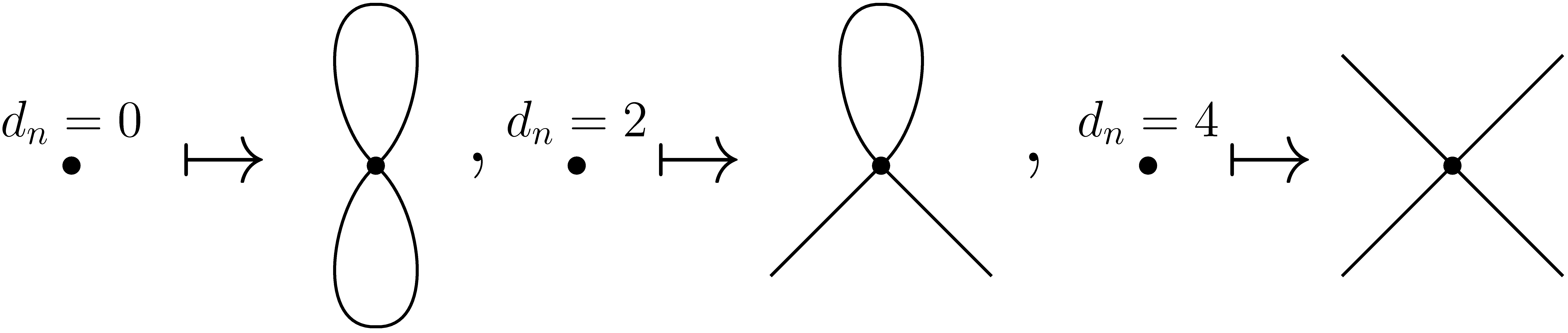}
	\caption{Correspondence between node labelling $d$ in $({\rm K}_4,d,m)$ and node structure in $\Gamma_{({\rm K}_4,d,m)}$.}
	\label{fig:node_notation}
\end{figure}
Alternatively one may read from $({\rm K}_4,d,m)$ the corresponding adjacency matrix. 
\begin{figure}[ht!]
	\centering
	\includegraphics[width=0.9\textwidth]{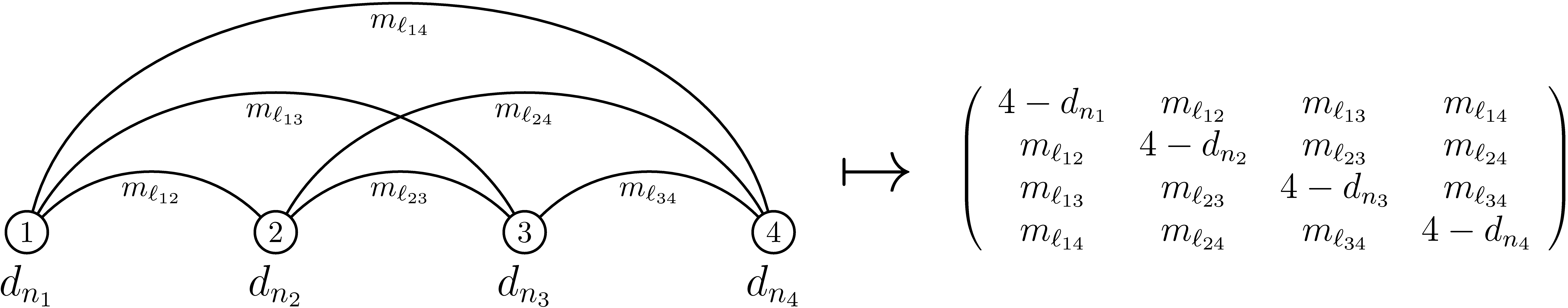}
	\caption{The adjacency matrix corresponding to $({\rm K}_4,d,m)$.}
	\label{fig:K4_adjacency}
\end{figure}
Simply choose some ordering of nodes (in our case it is a labelling of nodes with numbers $\{1,2,3,4\}$) and let $\ell_{ij}=\ell_{ji},\ i\not=j$ be the link in K${}_4$ connecting nodes $n_{i}$ and $n_{j}$. Now 
\begin{itemize}
	 \item terms $4-d_{n_i}$ correspond to diagonal entries of the adjacency matrix,i.e. $A_{\underline{i}\underline{i}}:=4-d_{n_i}$,
	 \item terms $m_{\ell_{ij}}$ correspond to off-diagonal entries, i.e. $A_{ij}:=m_{\ell_{ij}}$ for $i\not=j$.
\end{itemize}
This correspondence is depicted on \fref{K4_adjacency}. Note that, having known the numbers $d_n$, the total number of links going from nodes $n_1,\,n_2$ to nodes $n_3,\,n_4$ (we denote it by $k$) and the number $m_{\ell_{23}}$, we can reconstruct the remaining coloring of $({\rm K}_4,d,m)$. As a result those numbers give the parametrisation of adjacency matrix, we are using. Explicitly, we parametrize the solutions to equations \reef{czterowalencja} with
\begin{itemize}
	 \item four numbers $d_1,\,d_2,\,d_3,\,d_4 \in \{0,2,4\}$ satisfying triangle inequalities \reef{triangle_inequality},
	 \item a natural number $k\in \left[|d_1-d_2|, d_1+d_2\right]\cap\left[|d_3-d_4|,d_3+d_4\right]$, such that $k+d_1+d_2\in 2\mathbb{N}$,
	 \item an even natural number $m\in \left[d_3-d_4+d_2-d_1,{\rm min}\{d_3-d_4+k,d_2-d_1+k\}\right]$.
\end{itemize}
The corresponding parametrisation is:
\be
A=\left(\begin{array}{cccc} 4-d_{1}&\frac{d_1+d_2-k}{2}&\frac{d_2-d_1+k-m}{2}&\frac{d_4-d_3+d_1-d_2+m}{2}\\\frac{d_1+d_2-k}{2}&4-d_{2}&\frac{m}{2}&\frac{d_3-d_4+k-m}{2}\\\frac{d_2-d_1+k-m}{2}&\frac{m}{2}&4-d_{3}&\frac{d_3+d_4-k}{2}\\\frac{d_4-d_3+d_1-d_2+m}{2}&\frac{d_3-d_4+k-m}{2}&\frac{d_3+d_4-k}{2}&4-d_{4} \end{array}\right).
\ee
We next find orbits of action of permutation group $S_4$ on the set of those solutions. To this end we used Mathematica 8.0. The resulting graphs are depicted on \fref{interaction_graphs}.

\begin{figure}[ht!]
	\centering
	\includegraphics{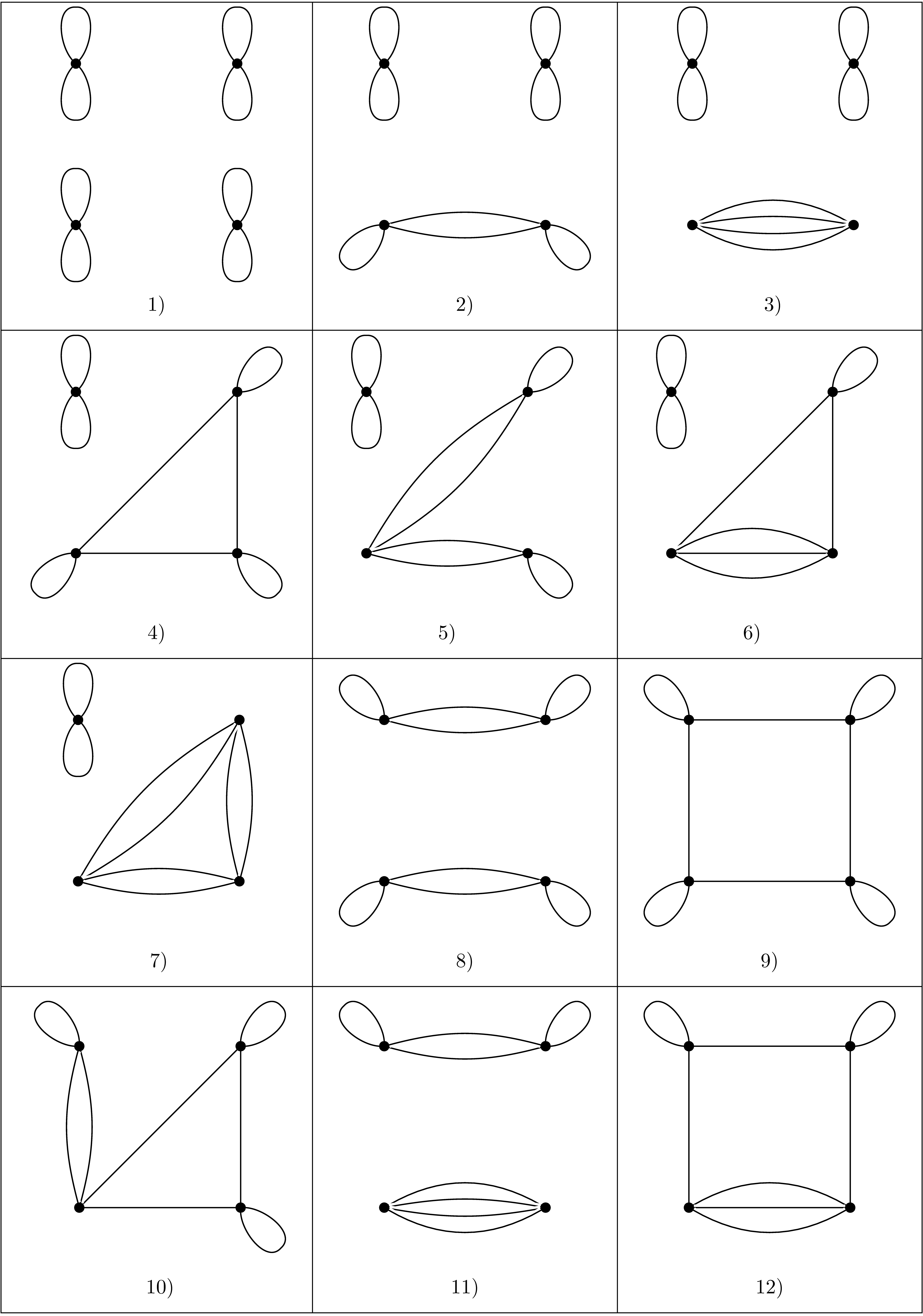}
\end{figure}
\begin{figure}[ht!]
	\centering
	\includegraphics{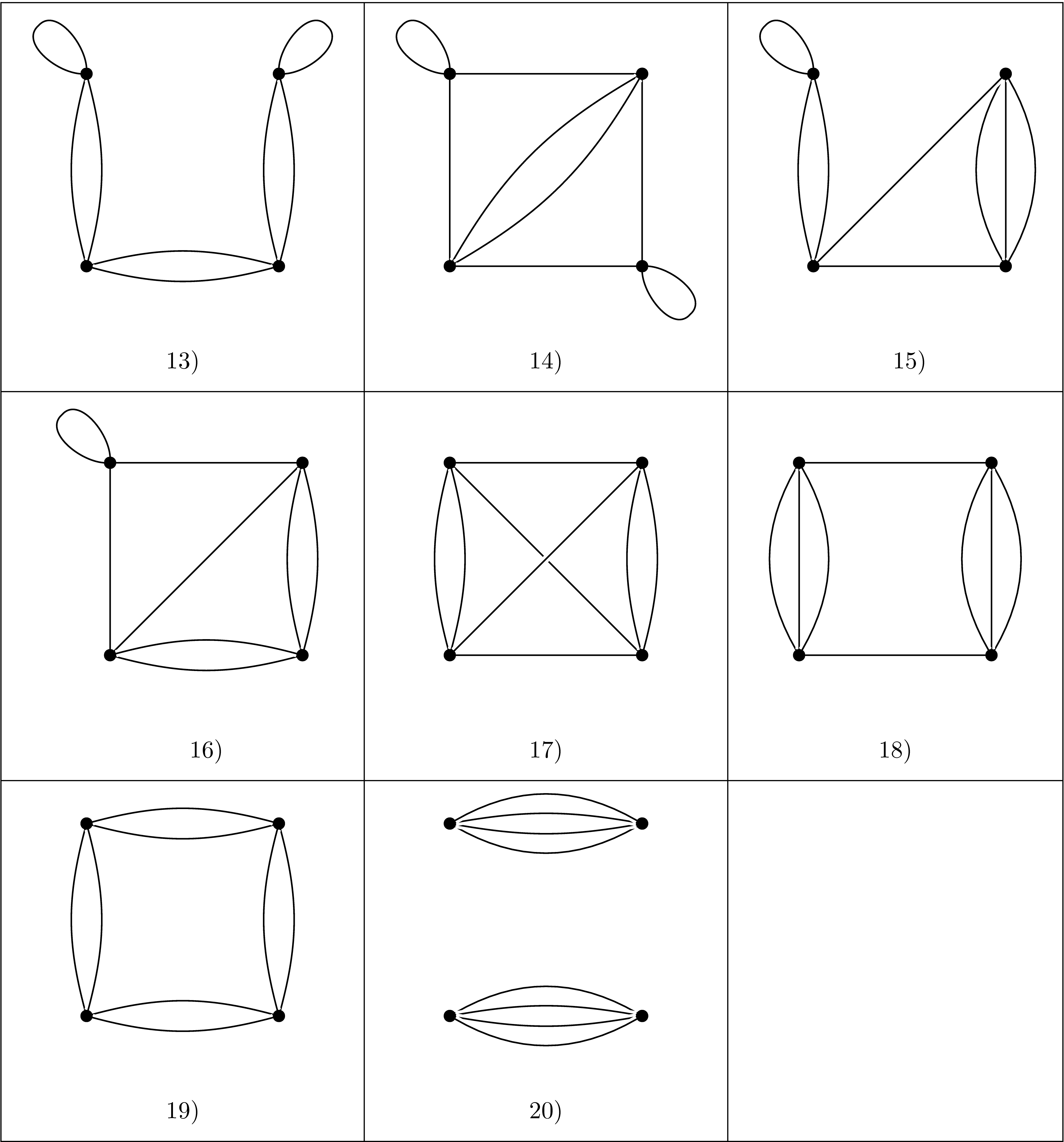}
	\caption{The list of all the possible interaction graphs  in the first order of the vertex expansion (modulo orientations).}
	\label{fig:interaction_graphs}
\end{figure}
 Note that one could further restrict the number of matrices considered by requiring that the sequence $(d_1,\,d_2,\,d_3,\,d_4)$ is monotonous and considering only orbits under action of $S_4/H$, where $H$ is the subgroup, which does not change the sequence $(d_1,\,d_2,\,d_3,\,d_4)$. This remark enables one to do the calculation without using computer. On the other hand, one could write a program which does not use the parametrisation we introduced -- e.g. one could generate matrices with entries taking values in the set $\{0,1,2,3,4\}$ (with even numbers on diagonal) and choose only those which satisfy equation \reef{czterowalencja} (a direct method). We have chosen the method we present here, because it gives better understanding of the structure of the graphs considered, it is less laborious than calculation by hand and the version we used is easier to implement than the direct method. It has the additional advantage that it is easily applicable to a more general case where the four nodes are not necessarily four-valent. When $d_1,\,d_2,\,d_3,\,d_4$ are becoming larger, this method becomes considerably faster than the direct method.

\subsection{Possible graph diagrams and an interesting observation}
As we explained in the previous  subsection, there are exactly 20 interaction graphs. However, the number of the graph diagrams resulting from the procedure described above is different. In this subsection we discuss in more details the diversity of the resulting graph diagrams.

Given an oriented interaction graph ${\cal D}_{\rm int}$ and a static diagram ${\cal D}_{\Gamma}$, there may be more than one graph diagrams ${\cal D}_{\Gamma}\#{\cal D}_{\rm int}$. The ambiguity is in the choice of the node relation and the link relations.
\begin{itemize}
	 \item {\bf The ambiguity in node relation.} It exists if an oriented interaction graph $\Gamma_{\rm int}$ has two nodes, say $n_1$ and $n_2$, such that the number of the incoming/outgoing links at $n_1$  is equal to the number of the incoming/outgoing links at $n_2$. Then, for every node relation between the nodes of the interaction graph and the corresponding nodes of the static diagram, there is another, different node relation obtained by switching the nodes $n_1$ and $n_2$ -- see \fref{node_relations}.
\begin{figure}[hbt!]
	\centering
	\subfloat[$\;$]{\label{fig:node_relation_1}\includegraphics[width=0.45\textwidth]{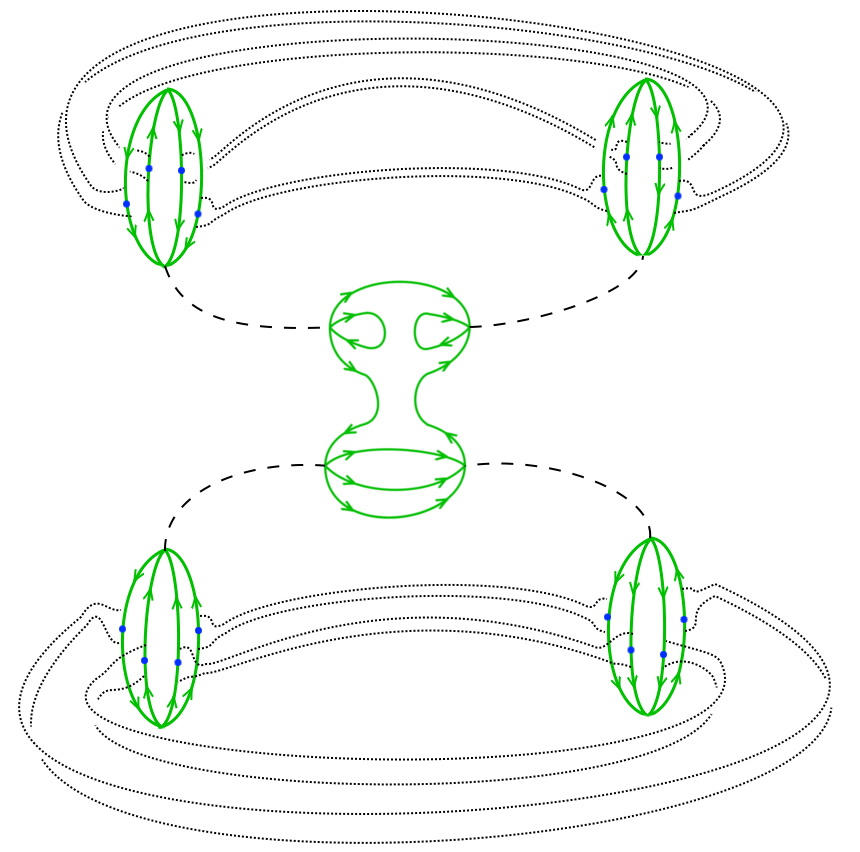}}
	\subfloat[$\;$]{\label{fig:node_relation_2}\includegraphics[width=0.45\textwidth]{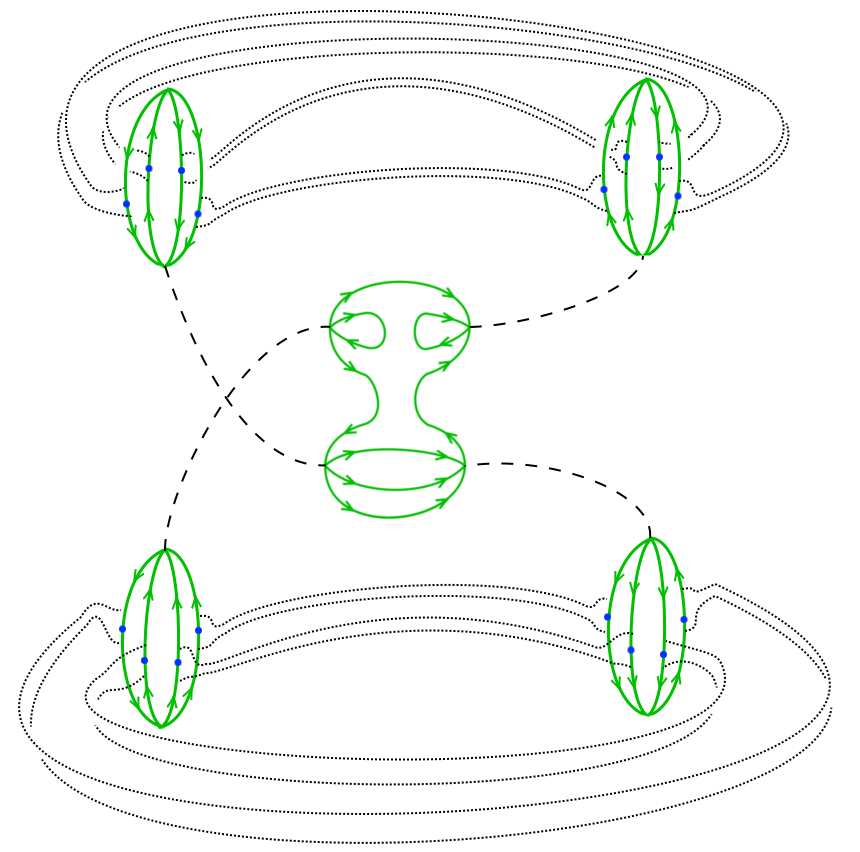}}
	\caption{Two nonequivalent graph diagrams ${\cal D}_{\Gamma}\#{\cal D}_{\rm int}$ obtained by different choices of a  node relation (the dashed lines) between the nodes of the interaction graph and the nodes in the static diagram.}
	\label{fig:node_relations}
\end{figure}
	 \item {\bf The ambiguity in  link relations.} Having settled down the node relation, there are still many possible link relations. The only condition, that each link relation needs to satisfy, is that incoming/outgoing link at each node in interaction graph is in relation with outgoing/incoming link at corresponding node in the static diagram (see \fref{link_relations}).
\begin{figure}[ht!]
	\centering
	\subfloat[$\;$]{\label{fig:link_relations_1}\includegraphics[width=0.45\textwidth]{3_graph_diagram}}
	\subfloat[$\;$]{\label{fig:link_relations_2}\includegraphics[width=0.45\textwidth]{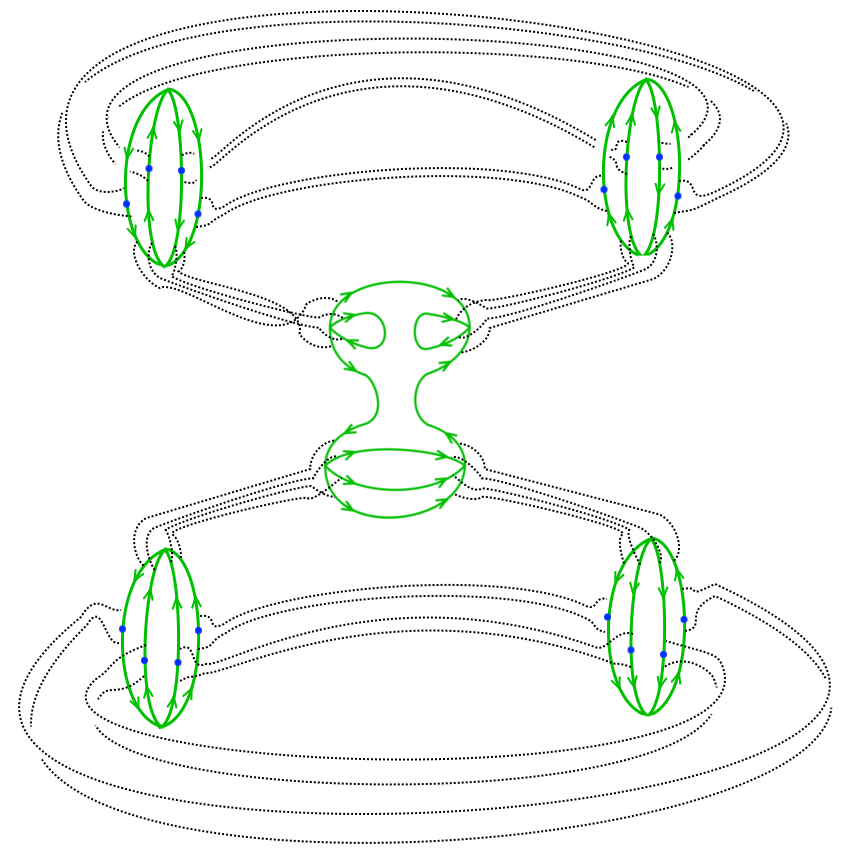}}
	\caption{Given an oriented interaction graph, a static diagram and a node relation one may choose different link relations between the links of the interaction graph and the corresponding links of the static diagram. Diagrams (a) and (b) are essentially different (the link relations are denoted by the dotted lines).}
	\label{fig:link_relations}
\end{figure}
\end{itemize}

Furthermore, \textbf{some colorings of the boundary links may be incompatible with some interaction graphs} --  it may happen that  the amplitude is zero for every coloring of a given interaction graph. In order to see how this limits the number of possible interaction graphs, consider the coloring of boundary graph depicted on \fref{3_thetas_colored} and OSN diagram on \fref{3_coloring} (node relations and the corresponding coloring with operators is omitted for clarity).  It is straightforward to see that the amplitude is non-zero only if the representations $\rho_2$ and $\rho_3$ and, respectively, the representations $\rho_1$ and $\rho_5$ are  equal ($\rho_2 = \rho_3$, $\rho_1 = \rho_5$). Importantly, note that, because there are links forming closed loops in the interaction graph, the amplitude is non-zero only if among representations $\rho_1,\,\rho_2,\,\rho_3,\,\rho_4$ or among representations $\rho_5,\,\rho_6,\,\rho_7,\,\rho_8$ there is a pair of equal representations. In addition, since the interaction graph is connected the amplitude is non-zero only if there is a pair of equal representations $\rho_i= \rho_j$, such that $i\in\{1,2,3,4\},\, j\in\{5,6,7,8\}$. Those two conditions do not depend on the choice of node and link relations but on the structure of interaction graph only. This example shows that there are colorings of boundary graph which are not compatible with the given interaction graph. 

\begin{figure}[ht!]
	\centering
	\subfloat[$\;$]{\label{fig:3_thetas_colored}\includegraphics[width=0.2\textwidth]{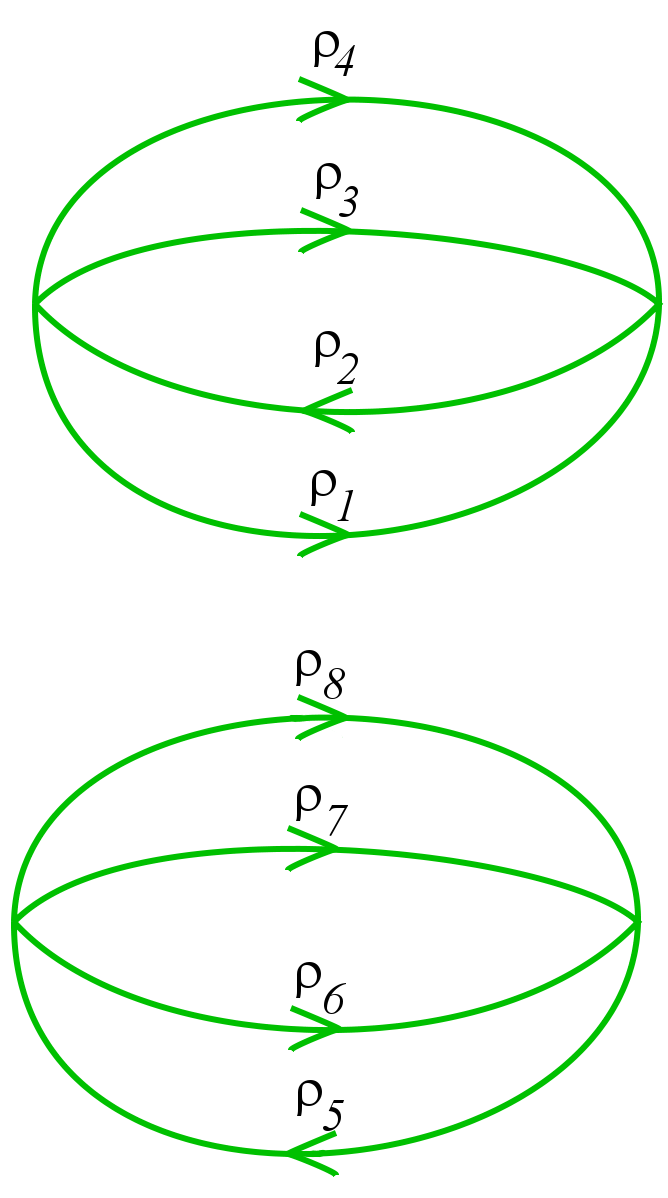}}\hspace{0.05\textwidth}
	\subfloat[$\;$]{\label{fig:3_coloring}\includegraphics[width=0.5\textwidth]{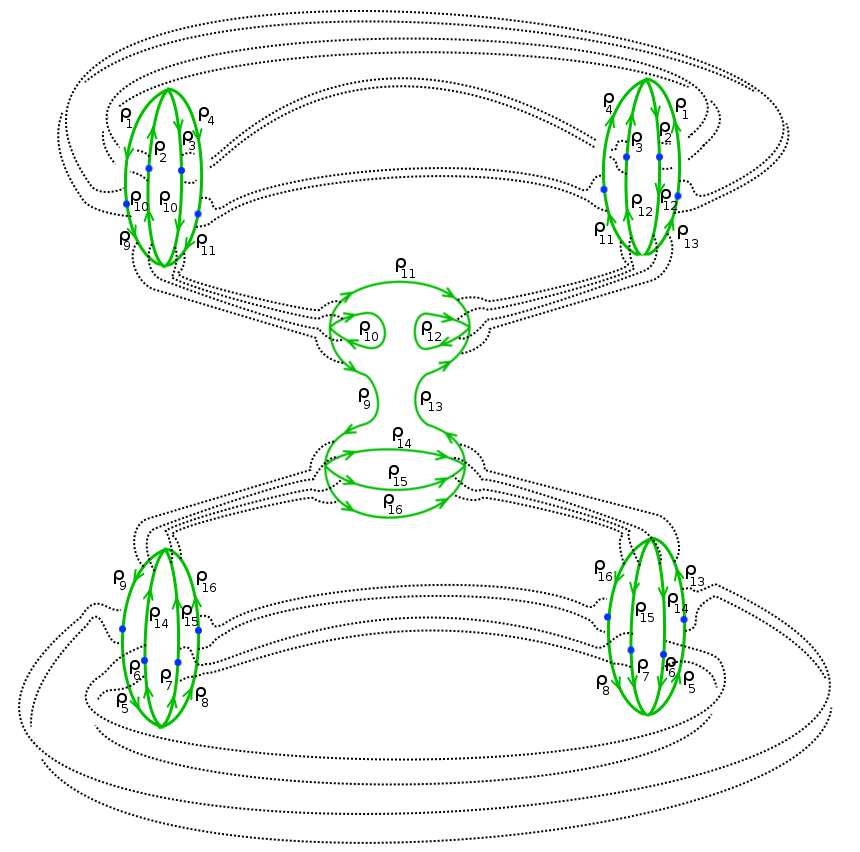}}
	\caption{Compatibility of a coloring of the boundary graph with a given interaction graph -- an example. (a) A coloring of the links of a boundary graph. (b) A coloring of the links  of a graph diagram. The link relations are denoted by dotted lines. The node relation and the corresponding coloring with operators are omitted for clarity of exposition. Note that the amplitude corresponding to OSN diagram from \fref{3_coloring} is non-zero only if $\rho_1$ is equal to $\rho_5$ and $\rho_2$ is equal to $\rho_3$.}
	\label{fig:coloring}
\end{figure}

A similar analysis may be performed for other interaction graphs. It leads to an interesting conclusion. There is a distinguished interaction graph, which is not limited by the coloring in the way described above -- it is the graph 20 from \fref{interaction_graphs} used in \cite{SF_cosmology}. The corresponding amplitude is non-zero even if all eight links of the boundary graph are labeled with pairwise different representations. In this generic case, all other interaction graphs give identically zero amplitude. We expect therefore that for a generic boundary state (which is a linear combination of  spin-network  states of all possible spins), the graph diagram with this interaction graph gives major contribution. This conclusion needs however further justification.

\StopkaPliku


\section{Summary, conclusions and outlook}
We presented a general algorithm for finding all spin-foams with given boundary graph in given order of vertex expansion. We applied this algorithm to the spin-foam cosmology model \cite{SF_cosmology} and found all contributions in first order of vertex expansion which are compatible with generalization of EPRL vertex \cite{SFLQG}. We expect that for a generic state the vertex used in \cite{SF_cosmology} gives the main contribution to the transition amplitude. This scenario needs however more thorough calculations and we leave it for further research.

The calculation we presented illustrates an application of OSN diagrams. The strength of this formalism lies in simplifying the classification of 2-complexes -- listing those with given properties (such as order of vertex expansion or structure of boundary graph). It also gives precise definition of the class of 2-complexes one should consider.

\section*{Acknowledgments}
Marcin Kisielowski and Jacek Puchta acknowledges financial support from the project "International PhD Studies in Fundamental Problems of  Quantum Gravity and Quantum Field Theory" of Foundation for Polish Science, cofinanced from the programme IE OP 2007-2013 within European Regional Development Fund. The work was also partially supported by the grants N N202 104838, and 182/N-QGG/2008/0 (PMN) of Polish Ministerstwo Nauki i Szkolnictwa Wy\.zszego. All the authors benefited from the travel grant of the ESF network Quantum Geometry and Quantum Gravity.

\appendix


\end{document}